\documentclass[acmsmall]{acmart}

\AtBeginDocument{%
  \providecommand\BibTeX{{%
    \normalfont B\kern-0.5em{\scshape i\kern-0.25em b}\kern-0.8em\TeX}}}

\setcopyright{acmcopyright}
\acmDOI{XXXXXXX.XXXXXXX}

\usepackage{multirow}
\usepackage{graphicx}
\usepackage{outlines}
\usepackage{tcolorbox}
\usepackage{wasysym}
\usepackage{colortbl}
\usepackage{soul}
\usepackage[normalem]{ulem}
\usepackage{soul}
\tcbuselibrary{breakable}
\usepackage[inline]{trackchanges}
\newtcbox{\xmybox}[1][red]{%
  on line,
  arc=7pt,
  colback=#1!10!white,
  colframe=#1!50!black,
  before upper={\rule[-3pt]{0pt}{10pt}},
  boxrule=1pt,
  boxsep=0pt,
  left=6pt,
  right=6pt,
  top=2pt,
  bottom=2pt 
}
\newtcbox{\mybluebox}[1][blue]{%
  on line,
  arc=7pt,
  colback=#1!10!white,
  colframe=#1!50!black,
  before upper={\rule[-3pt]{0pt}{10pt}},
  boxrule=1pt,
  boxsep=0pt,
  left=6pt,
  right=6pt,
  top=2pt,
  bottom=2pt 
}

\begin{document}

\title[The Power of AI in Qualitative Research]{Redefining qualitative analysis in the AI era: Utilizing ChatGPT for efficient thematic analysis}

\author{He Zhang}
\email{hpz5211@psu.edu}
\author{Chuhao Wu}
\email{cjw6297@psu.edu}
\author{Jingyi Xie}
\email{jzx5099@psu.edu}
\author{Yao lyu}
\email{yml5549@psu.edu}
\author{Jie Cai}
\email{jpc6982@psu.edu}
\author{John M. Carroll}
\authornote{Corresponding author.}
\email{jmc56@psu.edu}
\affiliation{%
  \institution{College of Information Sciences and Technology, Pennsylvania State University}
  \city{University Park}
  \state{PA}
  \country{USA}
  \postcode{16802}
}

\renewcommand{\shortauthors}{Zhang and Wu, et al.}

\begin{abstract}

AI tools, particularly large-scale language model (LLM) based applications such as ChatGPT, have the potential to simplify qualitative research. Through semi-structured interviews with seventeen participants, we identified challenges and concerns in integrating ChatGPT into the qualitative analysis process. Collaborating with thirteen qualitative researchers, we developed a framework for designing prompts to enhance the effectiveness of ChatGPT in thematic analysis. Our findings indicate that improving transparency, providing guidance on prompts, and strengthening users’ understanding of LLMs' capabilities significantly enhance the users' ability to interact with ChatGPT. We also discovered and revealed the reasons behind researchers' shift in attitude towards ChatGPT from negative to positive. This research not only highlights the importance of well-designed prompts in LLM applications but also offers reflections for qualitative researchers on the perception of AI’s role. Finally, we emphasize the potential ethical risks and the impact of constructing AI ethical expectations by researchers, particularly those who are novices, on future research and AI development.
\end{abstract}


\begin{teaserfigure}
  \includegraphics[width=1\columnwidth]{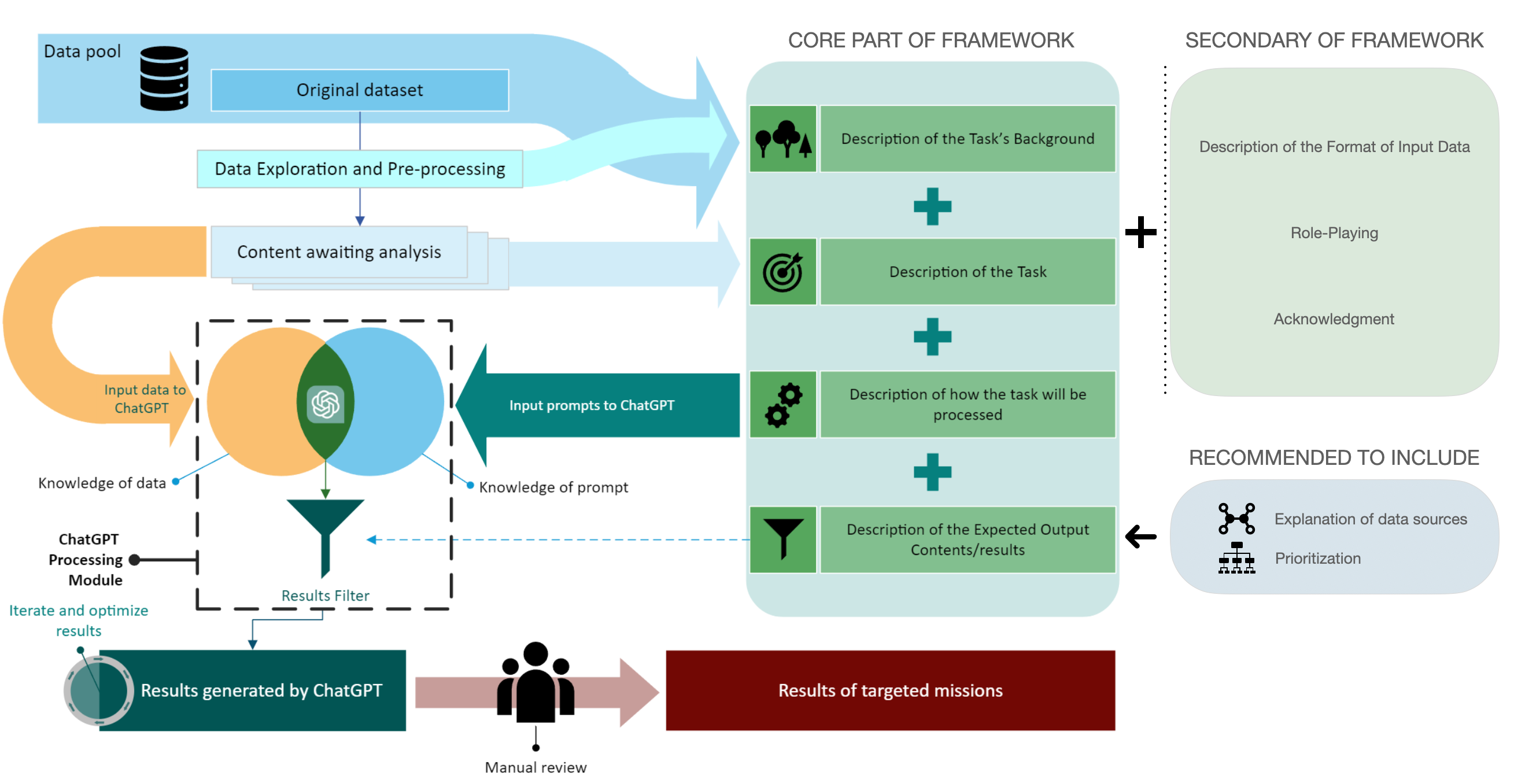}
  \caption{A workflow for applying ChatGPT to handle qualitative analysis tasks. The core of the (prompt design) framework includes descriptions of tasks (including methods), task backgrounds, and output format, enabling ChatGPT to analyze input data with strong robustness. The secondary part of the framework includes descriptions of data structure, role-playing, and friendly wording, which can further enhance the robustness of ChatGPT in task processing.}
  \label{fig.teaserfigure}
\end{teaserfigure}


\maketitle
\section{Introduction}
In the realm of qualitative research, thematic analysis is a highly flexible and widely used method for identifying, analyzing, and interpreting patterns of meaning (`themes') within qualitative data~\cite{braun2012thematic, braun2006using}. Despite its utility, thematic analysis can be time-consuming and require significant manual effort, especially when dealing with large and complex datasets~\cite{nowell2017thematic}. As we continue to generate increasingly massive volumes of qualitative data, there's an urgent need to seek out innovative methods to streamline and enhance the process of thematic analysis~\cite{bazeley2013qualitative}.

The realm of artificial intelligence (AI) is swiftly evolving, poised to revolutionize the ways we approach research~\cite{davenport2018artificial}. In particular, AI algorithms based on machine learning have showcased their prowess in efficiently processing and analyzing vast datasets~\cite{dwivedi2021artificial}. 

With the rise of Large-scale Language Models (LLMs), the capabilities of AI have been further enhanced. By 2023, ChatGPT stands out in this field, rooted in the ``conversational'' ability of ChatGPT allows users to interact with it through natural language, enabling non-AI expert users to use such AI for tasks more efficiently~\cite{10.1145/3544548.3580969}. It's distinguished by its exceptional ability to understand and generate human-like text, making this outstanding ability to understand and generate human-like text makes it a potential asset for qualitative researchers~\cite{radford2019language,10.1145/3641289}. Currently, ChatGPT makes human-AI interaction unprecedentedly direct and easy, requiring only textual prompts~\cite{NEURIPS2020_1457c0d6}. However, while entering prompts into LLMs seems effortless, designing effective prompts is not an easy task and may require additional experience~\cite{10.1145/3544548.3581388}, and the quality of the interaction results is highly related to the quality of the prompts~\cite{10.1145/3544548.3581388,fiannaca2023Programming}.

This study examines the potential of ChatGPT as an instrument for thematic analysis in qualitative research. We present a preliminary exploration into leveraging ChatGPT's capabilities, highlighting its prospective advantages and limitations, especially in the context of its practical implementation for thematic analysis. Consequently, we put forth the following research question (RQ):
\begin{itemize}
%
  \item[\textbf{RQ}.] \textbf{Can the performance of ChatGPT in qualitative analysis tasks be enhanced through prompt design? If so, how?} 
\end{itemize}

To address the research questions, we recruited 17 participants in total, and processed semi-structured interviews. We systematically analyzed these interviews to distill and summarize the challenges and concerns associated with using ChatGPT. Subsequently, we engaged 13 researchers of participants with expertise in qualitative research and analysis in the experiment. In the course of this collaborative project, we identified a range of techniques, user interactions, and conceptual approaches that significantly enhance ChatGPT's efficacy in qualitative analysis. These insights have been consolidated into a comprehensive prompt framework~(Fig.~\ref{fig.teaserfigure}), specifically designed to bolster ChatGPT’s capabilities in aiding the analysis of qualitative data. This framework stands out as a particularly timely and effective resource for those new to utilizing ChatGPT. Additionally, our research journey and its outcomes have catalyzed a series of pertinent discussions. These include: (1) examining the robustness and scope of ChatGPT's application; (2) exploring the potential roles of ChatGPT as either a co-researcher or a tool; (3) assessing the evolving ethical implications associated with the optimized use of such AI tools. Through this collaboration, we developed a set of cueing frameworks to facilitate qualitative data analysis with ChatGPT. Lastly, we delved into a discussion on the role of ChatGPT in qualitative analysis. The main content distribution, mind map, and workflow of the article are shown in Fig.~\ref{fig.teaserfigure}.
Through this exploration, we aspire to not only open up a new avenue for conducting thematic analysis but also contribute to the broader dialogue about the intersection of AI and qualitative research~\cite{bryman2016social}. The implications of this study are potentially far-reaching, affecting researchers, AI developers, and a wide range of professionals seeking to leverage AI in qualitative analysis~\cite{watkins2012qualitative}.

\section{Related Work}

\subsection{ChatGPT and Prompt Engineering}

ChatGPT, an advanced language model developed by OpenAI, has been recognized for its versatility across a wide range of language tasks, making it a powerful tool for various applications~\cite{10137850, teubner2023welcome}. Its capabilities include generating human-like text, content creation, sentence and paragraph completion, as well as essay and report writing~\cite{liebrenz2023generating,alkaissi2023artificial,bishop2023computer,macdonald2023can}. However, it is important to acknowledge ChatGPT's limitations~\cite{DWIVEDI2023102642}. The model can sometimes generate nonsensical or incorrect outputs, particularly when presented with ambiguous prompts~\cite{shen2023chatgpt,hassani2023role}.

The quality of output from LLMs like ChatGPT is significantly influenced by the instructions or ``prompts'' given to them~\cite{10.1145/3544548.3581388,fiannaca2023Programming}. Various studies have shown that specific prompt engineering techniques often yield more desirable outcomes. For instance, few-shot learning~\cite{NEURIPS2020_1457c0d6, zhao2021Calibrate} enables LLMs to learn and quickly produce a "new" model through prompt engineering, without any parameter updates. \citet{zhao2021Calibrate} demonstrated this by providing examples with emotional labels in their prompts, which led the LLM to more robustly perform emotion labeling tasks for other content using small but critical prompts, similar to a calibration process. \citet{10.1145/3491102.3501870} integrate natural language and code as prompt input, and the suggestions of \citet{10.1145/3544548.3581388} to add explanatory labels are similar to the logic of few-shot learning~\cite{zhao2021Calibrate}. Chain-of-thought approaches~\cite{wei2022ChainofThought} aim to reduce model errors by providing a series of thought samples, mirroring the human process of step-by-step reasoning. This method offers an interpretable insight into the model's analytical behavior, breaking down tasks into more detailed steps. \citet{10.1145/3543873.3587655} suggest adding prompts for understanding sentence structure and relationships, \citet{10.1145/3544548.3580969} explain the usage and content of prompts, and the prompt classification framework proposed by \citet{white2023prompt} essentially aims to establish a chain-of-thought in LLMs. Additionally, role-playing scenarios~\cite{gao2023Prompt} represent a straightforward strategy that can enhance LLMs' focus, encourage a specific creative style, and integrate the knowledge required by the "role" into the task.

While these related works indicate that the performance of ChatGPT can be enhanced through prompt engineering, they also highlight the challenges and limitations of current approaches. Effective utilization of ChatGPT varies across domains, and domain-specific knowledge is essential for optimizing its performance~\cite{tian2023Opportunities, wang2023Brief}. Expert advocate that prompt engineering should take the specific application context into account~\cite{heston2023Prompt}. However, there is a lack of research on how to design prompts for complex, open-ended tasks such as qualitative analysis, which require a deep understanding of the domain and the ability to interpret nuanced language. Moreover, existing prompt engineering techniques often focus on improving the accuracy and coherence of the generated output, but pay less attention to the transparency and explainability of the model's behavior, which are crucial for building trust and facilitating collaboration between humans and AI systems.

To address these gaps, our work explores a human-centric approach to prompt engineering for qualitative analysis with ChatGPT. We aim to develop a framework that empowers qualitative researchers, especially those new to the field, to effectively leverage ChatGPT's capabilities in their analysis workflows.

\subsection{Current Methodologies and Features of Thematic Analysis}
Thematic analysis~\cite{braun2012thematic} serves as a cornerstone in qualitative research, offering researchers a method to identify, analyze, interpret, and elucidate patterns or themes from their data. Its application transcends disciplines, contributing to its popularity and widespread use. Current practices encompass a range of methodologies~\cite{williams2019art,parameswaran2020live}, each bringing a unique perspective to the process of thematic analysis.

The Six-Phase Approach~\cite{braun2012thematic,clarke2015thematic,maguire2017doing} is widely recognized for its sequential process, guiding researchers through distinct phases, from data familiarization to final report preparation. This iterative methodology demands active researcher engagement in identifying themes. However, this engaged approach can be challenging for junior researchers who lack experience in qualitative analysis.

Boyatzis' Codebook Approach~\cite{boyatzis1998transforming} introduces a paradigm that necessitates creating a codebook before data coding. While beneficial for managing large datasets or collaborative coding~\cite{richards2018practical,guest2011applied}, developing a comprehensive codebook requires significant expertise, which may be daunting for novice researchers.

Thematic analysis poses several challenges, many of which are amplified for junior researchers. Subjectivity in identifying themes~\cite{morgan2022understanding} can lead to varying interpretations~\cite{10.1145/3544548.3580766}, resulting in multiple potential themes from the same dataset~\cite{10.1145/3491102.3517647}. This 'researcher subjectivity'~\cite{braun2019reflecting} necessitates reflexivity and transparency, which may be difficult for those new to the method.

Moreover, thematic analysis is often resource-intensive~\cite{10.1145/3479856}, particularly for large datasets~\cite{castleberry2018thematic,guest2013collecting}. The time and effort required for coding and identifying patterns can be overwhelming for junior researchers~\cite{terry2017thematic}. Replicability and generalizability are also significant concerns~\cite{leung2015validity,smith2018generalizability}, as the interpretive nature of the approach can lead to differing themes based on researcher perspectives~\cite{10.1145/3544548.3580766}.

In this study, we aim to explore how advanced AI tools can collaborate with and support junior researchers in conducting thematic analysis. By developing a human-centric prompt design framework for ChatGPT, we investigate how AI can be leveraged to address challenges related to subjectivity, resources, and replicability while empowering researchers to learn and apply thematic analysis effectively.

\subsection{AI-augmented Qualitative Analysis}
The increasing interest in Natural Language Processing (NLP) within the qualitative research community stems from its ability to analyze large volumes of text effectively. Sentiment analysis and topic modeling are the most frequently used NLP approaches for processing unstructured text data, such as patient feedback \cite{khanbhai2021applying}. However, these methods have limitations, including the constrained range of detectable sentiments in sentiment analysis \cite{yue2019survey} and the difficulty in interpreting topic modeling results \cite{churchill2022Evolution}. \citet{guetterman2018Augmenting} found that while NLP approaches may lack nuance compared to human coders, they can augment human efforts and improve efficiency. This aligns with the views of \citet{10.1145/3526113.3545681} and \citet{10.1145/3544548.3581352}, who suggest that AI should assist researchers in refining and evolving their coding rather than replace human analysts.

\citet{chen2018using} highlighted the fundamental differences between qualitative and quantitative methods as a major challenge in applying machine learning (ML) to qualitative coding, suggesting that ML support should focus on identifying ambiguity in coding. \citet{10.1145/3479856} discussed the relationship between AI and human analysts from the perspectives of collaboration, insight, and analysis, as well as AI's potential to foster criticality and reflectiveness in data handling. \citet{10.1145/3449168} showcased researchers' complex sentiments towards incorporating AI into qualitative analysis collaboration, particularly the allure of assistance and skepticism towards AI-driven analysis.

Recent attention has shifted towards using advanced AI models, such as LLMs, for qualitative analysis. \citet{katz2023Exploringa} demonstrated successful labeling of student comments using ChatGPT when provided with a label taxonomy. Comparisons between human and ChatGPT coding suggest that AI models may perform better in deductive rather than inductive analysis \cite{siiman2023Opportunities}. \citet{gao2023coaicoder} discovered that while AI as a mediator among human coders could accelerate collaborative coding, it may impact code diversity. Consequently, the quality of AI-assisted analysis remains a subject of debate \cite{10.1145/3544548.3581352}.

Interpretability and understandability play a key role in implementing ethical AI in practice \cite{10.1145/3599974}. Trust is crucial in human-AI collaboration, with perceived risk, AI capabilities, expectations, and user vulnerability shaping the degree of trust users place in AI \cite{10.1145/3442188.3445923}. Transparency, through explanations and communication about how AI works, is more effective in enhancing user trust than simply providing algorithmic interpretations \cite{zerilli2022transparency}.

Users prefer explanations that clarify decision-making processes and illustrate how specific actions influence outcomes \cite{10.1145/3579363}. XAI and Human-Centric Generative AI should cater to different users' needs, from AI experts requiring detailed visualizations to novices needing simpler explanations \cite{10.1145/3387166}. The effectiveness of transparency is context-dependent \cite{10.1145/3588594}, and starting with simplified models can help build foundational knowledge before gradually introducing complexity \cite{10.1145/3374218}. \citet{10.1145/3519264}'s conceptual framework categorizes user understanding support in intelligent systems into user mindsets, involvement, and knowledge outcomes, highlighting the multifaceted nature of user engagement with AI systems.

For LLMs like ChatGPT, natural language interfaces may enhance transparency by delivering effective explanations interactively \cite{10.1145/3579541}. Allowing users to make manual edits and visualize model decisions may further boost explainability and adaptability \cite{10.1145/3652028}. However, \citet{10.1145/3519266} noted that explanations are more beneficial when users have some level of domain expertise, suggesting that researchers should maintain strong data familiarity and verify reliability through cross-referencing when using AI for qualitative analysis \cite{christou2023Iow}.

ChatGPT, as an advanced AI tool, shows promise in advancing qualitative analysis but may also cause trust and ethical issues. This study aims to explore these aspects from the perspective of junior researchers, investigating how to leverage ChatGPT to support qualitative research while addressing potential negative issues. By examining the explainability and integration of ChatGPT in qualitative analysis, we assess the acceptance, concerns, trust, and ethical implications associated with using such models, and explore strategies to effectively leverage ChatGPT while mitigating potential negative impacts.

\section{Methods}

\subsection{Data Collection}
This study was conducted online through video conferencing software (e.g., Zoom). The research process was recorded for transcription and analysis with the informed and consent of the participants.

\subsubsection{Pilot Study}
\label{pilot_study}
To gain a deeper understanding of ChatGPT's capabilities, and to establish a foundation for guidelines for interviews and experiment, we initiated a pilot interview study involving four participants who have experience in using qualitative methods and ChatGPT. We recruited them via social media. All of the participants who took part in the pilot interview study had been involved in one or more projects using qualitative analytic methods within the past three years, and each of them had experience using ChatGPT. Among these participants, three had received formal training or education in qualitative data analysis. The aim of this pilot study was to investigate the challenges tied to ChatGPT usage. Each pilot interview study lasted for 1-1.5 hours.

The interviews were semi-structured. We began by introducing participants to the foundational concepts of qualitative analysis and probing their experiences in this realm. Following this, we explored the challenges they encountered both in the broader scope of qualitative analysis and specifically when using ChatGPT. The thematic analysis centered on users' reflections — encompassing uses, challenges, and strategies — during their interactions with ChatGPT. Through a thorough review, analysis, and reflection on the recorded sessions and their respective coding, we found that participants indeed have some concerns and challenges in using ChatGPT, but at the same time, they also demonstrated enthusiasm for ChatGPT. This is mainly due to its rapid data processing and improved work efficiency (P1, P2, P3, P4), providing concise overviews or summaries (P2, P3, P4), generating preliminary insights (P1, P2, P3, P4), and its user-friendly question-and-answer interaction format (P4).
In addition, all participants in the pilot interviews study expressed interest in understanding how to use ChatGPT better and how to design prompts more effectively.

Based on these feedback, we refined the interview guide and developed an formal study's protocol to further explore ChatGPT's capabilities, uses, and strategies for addressing challenges in qualitative research. In the lens of formal study and coupled with qualitative scenario analysis, we devoted special attention to these challenges, with an aspiration to pinpoint potential solutions for RQ.


\subsubsection{Formal Study}
\label{formal_study}
We paid particular attention in this section to how participants' design of the prompts affected ChatGPT's performance, their strategies in qualitative analysis, and expected outcomes. We worked with participants through formal study to design ChatGPT prompts that were friendly to qualitative analysis~\cite{levac2019scoping}. Finally, we distilled from the design solutions a framework of prompts to be applied in ChatGPT for qualitative analysis tasks.

The formal study comprised two parts: interviews and experiment. All researchers convened for two meetings to develop the protocol, make consensus on the procedure, and conduct internal testings prior to the formal study.

The interviews were semi-structured, focusing primarily on two areas: (1) challenges, approaches, and techniques encountered in qualitative analysis research, and (2) experiences, challenges, and insights related to using ChatGPT. Additionally, for participants who had experience with qualitative analysis software (e.g., NVivo and atlas.ti), we inquired about their experiences, benefits, and limitations with these tools. This part took approximately 20 minutes.

Next, we sent participants a corpus of qualitative data (a transcript of a focus groups), and asked participants to use ChatGPT to perform qualitative analysis coding on the content in the corpus while sharing screen. All participants interacted with ChatGPT in English. First, we allowed participants to design their own prompts (up to 5 times). During this process, researchers asked participants to think aloud and provide a detailed explanation of the intentions, requirements, and strategies in each designed prompt, and to evaluate the generated results. During the iteration (updating prompts) process, we also asked participants to explain their intentions and provide reasons for modifications. In addition, we asked participants to compare and evaluate the generated results from previous interactions and provide reasons for their respective attitudes. 

Second, the researchers joined the prompt design process and made suggestions on the prompts that the participants had previously designed independently. Specifically, researchers and participants discussed the outcomes of using the independently designed prompts and collaboratively refined them to achieve better results. In this phase, researchers integrated their own experience and insights from previous work, including few-shot learning~\cite{zhao2021Calibrate}, chain-of-thought approaches~\cite{wei2022ChainofThought}, role-playing~\cite{gao2023Prompt}, adding explanatory tags~\cite{10.1145/3544548.3581388}, categorizing prompts~\cite{white2023prompt}, understanding sentence structure and relationships~\cite{10.1145/3543873.3587655}, considering instruction usage and content~\cite{10.1145/3544548.3580969}, and the integration of natural language and code in mixed inputs~\cite{10.1145/3491102.3501870} into the design of new prompts. Researchers also inquired about the participants' satisfaction with the newly generated results and their optimization suggestions. This was a continuous trial-and-error process, with researchers and participants co-designing prompts until the generated results met the participants' expectations. In this part, researchers and participants discussed together the generated results and the design rationale of the prompts.
Afterwards, researchers collaborated with the participants to review the prompts from this process that helped enhance the performance of ChatGPT in processing qualitative data. 
%





\newtcolorbox{mybox}{colback=blue!5!white,colframe = blue!75!black}

\subsection{Participants Recruitment}
We recruited 17 participants (12 females and 5 males) through social media and the authors' network. The age of the participants ranged from 20 to 32 years old (median = 27, SD~$\approx$~3.70). P1-P4 were recruited to take part in the pilot study (Section~\ref{pilot_study}), while the remaining 13 participated in the formal study (Section~\ref{formal_study}). Detailed demographic information is presented in Table~\ref{tab:demographics}.
 Except for one participant, in the pilot study, all others had undergone formal training or courses about qualitative analysis. For formal study, we reached out to potential participants via our academic network, most of whom hailed from universities and were actively involved in qualitative research, while being fluent in English (if they are not native speakers). All 13 participants were researchers skilled in qualitative methods and had prior experience with ChatGPT. Among these participants, 11 were doctoral students, one was a master's student, and 1 worked in the industry but held a doctoral degree. All participants had formal training or education in qualitative data analysis. The pilot and formal studies were processed under approval of university IRB. Informed consent was obtained from all participants. All participants received \$10 gift card (or equivalent) for their time and effort.

\begin{table}[htbp]
\centering
\resizebox{\columnwidth}{!}{%
\begin{tabular}{ccrrrrcr}
\hline
Index & Age & Gender & Major & Occupation & Region & \begin{tabular}[c]{@{}c@{}}Training in \\ Qualitative \\ Analysis\end{tabular} & \begin{tabular}[c]{@{}r@{}}Knowledge of \\ Qualitative\\ Analysis\end{tabular} \\ \hline
\multicolumn{8}{c}{Pilot study} \\ \hline
1 & 25-30 & Male & Healthcare & Industry & East Asia & $\checkmark$ & Passing Knowledge \\
2 & 31-35 & Male & IT Consulting Services & Industry & East Asia & $\checkmark$ & Knowledgeable \\
3 & 18-24 & Female & Enological Engineering & Undergraduate Student - Junior & East Asia & $\times$ & Passing Knowledge \\
4 & 25-30 & Female & Electronic Engineering & Graduate Student - PhD & East Asia & $\checkmark$ & Passing Knowledge \\ \hline
\multicolumn{8}{c}{Formal Study} \\ \hline
5 & 25-30 & Female & Human-Computer Interaction & Graduate Student - PhD & North America & $\checkmark$ & Knowledgeable \\
6 & 25-30 & Female & Communication & Graduate Student - PhD & North America & $\checkmark$ & Knowledgeable* \\
7 & 18-24 & Female & Design & Graduate Student - Master & East Asia & $\checkmark$ & Passing Knowledge \\
8 & 25-30 & Female & Human-Computer Interaction & Graduate Student - PhD & East Asia & $\checkmark$ & Knowledgeable \\
9 & 25-30 & Male & Information Science & Graduate Student - PhD & North America & $\checkmark$ & Knowledgeable \\
10 & 25-30 & Female & Learning, Design and Technology & Graduate Student - PhD & North America & $\checkmark$ & Knowledgeable \\
11 & 18-24 & Male & Computer Science & Graduate Student - PhD & East Asia & $\checkmark$ & Knowledgeable \\
12 & 25-30 & Female & Human-Computer Interaction & Graduate Student - PhD & North America & $\checkmark$ & Knowledgeable \\
13 & 31-35 & Female & Human-Computer Interaction & Graduate Student - PhD & North America & $\checkmark$ & Knowledgeable \\
14 & 25-30 & Female & Information Science & Industry - PhD & North America & $\checkmark$ & Expert \\
15 & 25-30 & Male & Human-Computer Interaction & Graduate Student - PhD & North America & $\checkmark$ & Knowledgeable \\
16 & 25-30 & Female & Information Science & Graduate Student - PhD & North America & $\checkmark$ & Knowledgeable \\
17 & 25-30 & Female & History & Graduate Student - PhD & North America & $\checkmark$ & Knowledgeable \\\hline
\multicolumn{8}{l}{* Researchers who specialize in the prompt design.}
\end{tabular}%
}
\caption{Demographics}
\label{tab:demographics}
\end{table}



\subsection{Data Analysis}
The researchers conducted reflexive thematic analysis (RTA) on the collected data~\cite{clarke2015thematic, 10.1145/3491102.3517716}. The data analysis generally adhered to a six-step procedure: dataset familiarization, data coding, initial theme generation, theme development and review, theme refinement and definition, and report composition. After each interview and experiment, the research team briefly discussed the outcomes. All recordings were transcribed and coded by the first author and at least one other author. During this research process, the researchers met at least once a week to discuss the progress of the study, the results of the interviews and experiments, the findings and problems, and to continually refine the themes and processes.

\section{Users' Experiences and Challenges with ChatGPT}
In this section, we summarized participants' concerns when using ChatGPT to assist with qualitative analysis tasks. Specifically, participants expressed significant concerns about ChatGPT's transparency (interpretability and verifiability), performance (consistency and accuracy), the difficulty of designing prompts, and the cost of reviewing results. These findings further guided our subsequent design efforts.

\subsection{Lack of Transparency}

Participants' initial skepticism towards using ChatGPT for qualitative analysis was largely rooted in concerns about the lack of transparency in its outputs. As P13 mentioned:

\begin{quote}
``\textit{It's so convenient, both convenient and scary. Although it's convenient and might be smarter than me, I still don't know who it is, because I can't identify it just using keywords.}'' (P13)
\end{quote}
This highlights the unease researchers felt about relying on an AI tool without a clear understanding of how it generated its responses.

\subsection{Difficulty of Designing Propmpts}

Participants reported that the current cost of designing prompts and learning how to design prompts is high. The internet is flooded with an excess of prompt design schemes and tutorials, often leaving users confused about where to start. As P8 mentioned, ``\textit{There are too many junk prompts [tutorials/examples] online right now.}''

Additionally, when users engage in designing prompts, it frequently necessitates multiple attempts, the outcomes of which appear to be random. Also, the efficiency and accuracy with which ChatGPT operates is inextricably linked to the caliber of the prompts it receives. On one hand, ChatGPT offers the allure of automated and efficient outputs, while on the other, the effort to finely tune and optimize these prompts often verges on the boundary of practicality. A balance needs to be struck between automated ease and manual customization. This sentiment is encapsulated in the following quotes:

\begin{quote}
``\textit{I need to test it for a while. For instance, using ChatGPT, various prompts might lead to divergent outcomes, and I might have to repeatedly test to determine which type of prompt elicits the desired result.}'' (P14)
\end{quote}
\begin{quote}
``\textit{The process (of designing prompts) is also a bit time-consuming, and it might be easier for me to do some coding work manually.}'' (P8)
\end{quote}
This sentiment underscores the inherent dilemma faced by many developers and users.  The high overhead of designing effective prompts may cause potential users to default to their manual processes, thus underscoring the pressing need for a more streamlined and intuitive prompt-design framework.

\subsection{Insufficient Understanding of ChatGPT's Capabilities}

Participants' lack of knowledge about ChatGPT's capabilities or incorrect usage can lead to a decline in performance and discourage users from using the tool. As previously mentioned, the learning cost associated with understanding ChatGPT's capabilities is substantial in the context of information overload. Participants in our study were neither expert users of ChatGPT nor researchers in related applications, and their understanding of its capabilities was limited.

\begin{quote}
``\textit{I didn't know it could generate tables before.}'' (P17) 
\end{quote}
\begin{quote}
``\textit{When I used it before, I didn't know what I wanted... I didn't know what ChatGPT could do, so I couldn't maximize its performance. Using ChatGPT still feels like using a Swiss Army knife for work, not very efficient.}'' (P15). 
\end{quote}
\begin{quote}
``\textit{I'm quite shocked. I felt I didn't expect it to have these capabilities, because I hadn't tried asking in so many ways before.}'' (P12)
\end{quote}
\begin{quote}
``\textit{At first, I didn't expect ChatGPT to generate such a result for me, but I find it quite aligned with the ideal result.}'' (P7)
\end{quote}

The feedback from participants directly reflects the awkward situation of limited use scenarios due to a lack of understanding of ChatGPT's capabilities. P15 used a current example to analogize this awkward situation, indicating that even when using the ``most powerful'' AI systems to date, they can only be effectively applied to tasks if the users understand how to utilize them.

\subsection{Demand for Customized Solutions}

Although our findings emphasize the importance of being informed about ChatGPT's capabilities, this does not mean that users need to understand all of its functions completely. In other words, what users more urgently need is knowledge of how to use ChatGPT to better accomplish the tasks they are truly concerned with. This was manifested in our study as a demand for customized solutions. After introducing some of ChatGPT's capabilities and suggestions on how to use it for processing qualitative data to the participants, they provided positive feedback. This does not necessarily indicate their recognition of ChatGPT's abilities in other domains, but it did increase their understanding of its capabilities in handling qualitative data. Some illustrative quotes are as follows:
\begin{quote}
``\textit{I think especially for these prompts. You know, each [task's] topic might be different, like the prompts for programming and the prompts for qualitative research, they might all be different. I think at least for my future qualitative research, I will definitely follow the steps you suggested for writing prompts. They are indeed more specific than what I used to write.}'' (P10)
\end{quote}

Furthermore, P7 highlighted the need for a standardized yet flexible framework, offering clear, concise guidance on input formats and expected outcomes. This approach would streamline prompt design, ensuring efficiency in interactions with AI tools like ChatGPT and potential higher quality in outcomes.

\begin{quote}
``\textit{I think there can be a more standardized [process] because everyone has a procedure when doing qualitative analysis... Maybe the software tutorial will tell me what to input or in what format to achieve a certain result. Just present these processes to me. There's no need to be overly detailed...}'' (P7)
\end{quote}

The findings discussed above not only highlight the challenges participants faced when using ChatGPT for qualitative research analysis but also emphasize the need for a standardized yet flexible framework that provides suggestions and guidance on effectively utilizing ChatGPT. Participants' demands for customized solutions and clear guidance on input formats and expected outcomes underscore the importance of developing a framework that addresses these concerns. To bridge this gap and create a framework that meets the needs of qualitative researchers, we further collaborated with participants and engaged in an iterative analysis and design process. This process, detailed in the following section, aimed to develop a framework that provides structured guidance while maintaining flexibility, ultimately enabling researchers to harness the power of ChatGPT for their qualitative analysis tasks effectively.

\section{Analyses of the Design Process}\label{sec:promptdesignprocess}

Prompts serve as a pivotal mechanism in shaping the discourse of AI-human interaction, particularly with LLMs such as ChatGPT~\cite{10.1145/3544548.3581388,fiannaca2023Programming}. These initial inputs, or ``prompts'', can markedly impact the quality, coherence, and applicability of the language model's subsequent responses. It is not only an advantage, but also a challenge for applications of LLMs to accurately express intentions through natural language.

Inputting structured prompts is crucial for maintaining high performance in ChatGPT~\cite{jiang2023structgpt,doi:10.1080/08982112.2023.2206479}. Although the phrasing of the prompts can vary, they should contain fundamental pieces of information that form a core element of the pattern~\cite{white2023prompt}. Hence, we summarize the key elements/features/rules/strategies possessed by the prompts used in achieving the participants' desired (satisfactory) outcomes, as shown in Table~\ref{tab:summaryprompts}. This includes strategies in the independent coding process of participants, as well as various prompt design methods and testing strategies proposed by researchers based on the goals expected by the participants, which are considered effective by the participants. The specific process to determine these features was as follows: (1) We analyzed the results generated by ChatGPT for each participant during the formal study. (2) We selected the output results that participants deemed satisfactory, those that participants believed could be used directly as results or had completed some important qualitative analysis tasks for them, such as thematic analysis. (3) We reviewed the prompts that led to the corresponding output results and conducted a comprehensive analysis in conjunction with the interview content to determine the specific feature categories.

\begin{table}[htbp]
\resizebox{\columnwidth}{!}{%
\begin{tabular}{c|ccccccccc}\hline
\begin{tabular}[c]{@{}c@{}}Participant \\ No.\\ (Index)\end{tabular} & \begin{tabular}[c]{@{}c@{}}Background /\\ Conceptual \\ Understanding\end{tabular} & \begin{tabular}[c]{@{}c@{}}Focus \\ on \\ Methodology\\ (Goal of Task)\end{tabular} & \begin{tabular}[c]{@{}c@{}}Focus on \\ Analytical \\ Process (context)\end{tabular} & \begin{tabular}[c]{@{}c@{}}Data Format\\ (Inputs)\end{tabular} & \begin{tabular}[c]{@{}c@{}}Data Format\\ (Outputs)\end{tabular} & Role-Playing & Prioritization & \begin{tabular}[c]{@{}c@{}}Transparency \\ \& \\ Traceability\end{tabular} & \begin{tabular}[c]{@{}c@{}}Acknowledgment \\ of \\ Expertise\end{tabular} \\ \hline
5 & $\CIRCLE$ & $\CIRCLE$ & $\CIRCLE$  & $\Circle$ & $\Circle$ & $\Circle$ & $\CIRCLE$ & $\CIRCLE$ & $\Circle$ \\
\rowcolor[HTML]{EFEFEF} 
6 & $\CIRCLE$ & $\CIRCLE$ & $\Circle$ & $\Circle$ & $\Circle$ & $\CIRCLE$ & $\CIRCLE$  &  $\CIRCLE$ & $\Circle$ \\
7 & $\CIRCLE$ & $\CIRCLE$ & $\CIRCLE$ & $\Circle$ & $\Circle$ & $\CIRCLE$ & $\CIRCLE$  &  $\CIRCLE$ & $\Circle$ \\
\rowcolor[HTML]{EFEFEF} 
8 & $\CIRCLE$ & $\CIRCLE$ & $\Circle$ & $\RIGHTcircle$ & $\RIGHTcircle$ & $\CIRCLE$ & $\Circle$  &  $\CIRCLE$ & $\CIRCLE$ \\
9 & $\CIRCLE$ & $\CIRCLE$ & $\Circle$ & $\RIGHTcircle$ & $\RIGHTcircle$ & $\Circle$ & $\CIRCLE$  &  $\CIRCLE$ & $\Circle$ \\
\rowcolor[HTML]{EFEFEF} 
10 & $\CIRCLE$ & $\CIRCLE$ & $\CIRCLE$ & $\Circle$ & $\Circle$ & $\Circle$ & $\CIRCLE$ &  $\CIRCLE$ & $\Circle$ \\
11 & $\CIRCLE$ & $\CIRCLE$ & $\CIRCLE$ & $\Circle$ & $\RIGHTcircle$ & $\Circle$ & $\CIRCLE$ &  $\CIRCLE$ & $\Circle$ \\
\rowcolor[HTML]{EFEFEF} 
12 & $\CIRCLE$ & $\CIRCLE$ & $\CIRCLE$ & $\RIGHTcircle$ & $\RIGHTcircle$ & $\Circle$ & $\Circle$ &  $\CIRCLE$ & $\Circle$ \\
13 & $\CIRCLE$ & $\CIRCLE$ & $\CIRCLE$ & $\Circle$ & $\Circle$ & $\CIRCLE$ & $\CIRCLE$ &  $\CIRCLE$ & $\Circle$ \\
\rowcolor[HTML]{EFEFEF} 
14 & $\CIRCLE$ & $\CIRCLE$ & $\CIRCLE$ & $\RIGHTcircle$ & $\RIGHTcircle$ & $\Circle$ & $\CIRCLE$ &  $\CIRCLE$ & $\Circle$ \\
15 & $\CIRCLE$ & $\CIRCLE$ & $\CIRCLE$ & $\RIGHTcircle$ & $\RIGHTcircle$ & $\Circle$ & $\CIRCLE$ &  $\CIRCLE$ & $\Circle$ \\
\rowcolor[HTML]{EFEFEF} 
16 & $\CIRCLE$ & $\CIRCLE$ & $\CIRCLE$ & $\Circle$ & $\Circle$ & $\Circle$ & $\Circle$ &  $\CIRCLE$ & $\Circle$ \\
17 & $\CIRCLE$ & $\CIRCLE$ & $\Circle$ & $\Circle$ & $\RIGHTcircle$ & $\CIRCLE$ & $\Circle$ &  $\CIRCLE$ & $\CIRCLE$ \\ \hline
\multicolumn{10}{l}{$\Circle$: a strategy not employed by the participant in crafting the prompt.}\\
\multicolumn{10}{l}{$\CIRCLE$: a strategy utilized by the participant in crafting the prompt.}\\
\multicolumn{10}{l}{$\RIGHTcircle$: a strategy adopted by the participant following the researcher's suggestion.}

\end{tabular}%
}
\caption{Summary of Participant Design Prompts}
\label{tab:summaryprompts}
\end{table}

\begin{table}[htbp]
\centering
\resizebox{\columnwidth}{!}{%
\begin{tabular}{ll}
\hline
Strategies & Examples of Prompts \\ \hline
Background / Conceptual Understanding & \begin{tabular}[c]{@{}l@{}}``\textit{Here is a transcript from a focus group interview about 'Transitioning to Remote Work'.} \\ \textit{Each paragraph is from one participant. Please read it first.}'' (P9)\end{tabular} \\
\rowcolor[HTML]{EFEFEF} 
Focus on Methodology (Goal of Task) & ``\textit{can you do a thematic analysis of their responses?}'' (P6) \\
Focus on Analytical Process (context) & ``\textit{Please do a thematic analysis and summarize no more than 10 themes from this transcript.}'' (P9) \\
\rowcolor[HTML]{EFEFEF} 
Data Format (Inputs) & \begin{tabular}[c]{@{}l@{}}``\textit{The format of the transcript looks like this: Participant \textless{}participant name\textgreater :} \\ \textit{\textless{}transcript of this participant's comments regarding transitioning to remote work\textgreater{}.}'' (P15)\end{tabular} \\
Data Format (Outputs) & \begin{tabular}[c]{@{}l@{}}``\textit{Please analysis data again and make outputs follow format as below: }\\ \textit{(New line) Topic: \{challenges\}, \{strategies\}, \{benefits\}, \{downsides\}, \{productivity\}}\\ \textit{(New line) Argument points: \{reasons\}, \{suggestions\}, \{perceptions\}}\\ \textit{(New line) Supports: \{raw data\}, \{participant number\}}'' (P8)\end{tabular} \\
\rowcolor[HTML]{EFEFEF} 
Role-Playing & ``\textit{You are now a research expert in qualitative analysis...}'' (P7) \\
Prioritization & \begin{tabular}[c]{@{}l@{}}``\textit{I want a codebook with less than 10 codes. I hope these codes are main themes from }\\\textit{the transcript ...}'' (P10)\end{tabular} \\
\rowcolor[HTML]{EFEFEF} 
Transparency \&Traceability & ``\textit{...and tell me what’s your rationale for your analysis.}'' (P13) \\
Acknowledgment of Expertise & \begin{tabular}[c]{@{}l@{}}``\textit{Good job – I cannot finish my work without you, you are so sweet. Follow the requirements as below,} \\ \textit{and continue analyzing data.}'' (P5)\end{tabular} \\ \hline
\end{tabular}%
}
\caption{Strategies and examples.}
\label{tab:promptsexample}
\end{table}
Please note that the strategies presented in Tables~\ref{tab:summaryprompts} and~\ref{tab:promptsexample} are the results of summaries, and the prompts used by participants were not exactly the same. Besides, there are subtle interconnections between these strategies, which we will explain in more detail when describing these strategies in the following sections.

\subsection{Explanation of Prompts Provided}\label{sec:explanation-of-prompts-provided}

We will elaborate on the characteristics and reasons that should be considered in prompt design in this section, where the prompts used in formal study will be shown in Table~\ref{tab:summaryprompts} and~\ref{tab:promptsexample}.
\subsubsection{\textbf{Prompts Provided: Background or Conceptual Understanding}}
Within the prompts, utilizing descriptive language to offer a task background to ChatGPT is necessary. This approach equips the AI model with context, enabling a more nuanced understanding of the task at hand. By providing a detailed description of the task, including its purpose, the expected outcomes, and any specific nuances or considerations that should be taken into account, we help guide the AI in generating appropriate and targeted responses. This method enhances the quality of the generated outputs and increases the likelihood of successful data analysis. 

\subsubsection{\textbf{Prompts Provided: Focus on Methodology (Goal of Task)}}
The prompt should clearly and detailedly specify the specific task that ChatGPT should perform. For instance, in this study, the task of the AI model was to support the analysis of qualitative data related to the theme of 'remote work.' All participants clearly described this task in the prompt, pointing out that ChatGPT should analyze the given text data to find patterns and themes. Moreover, through testing, we found that using a more specific task improves ChatGPT's performance in handling the task. The level of detail in the task description often depends on the participant's experience, as P6 said: 
\begin{quote}
``\textit{Because I am familiar with things like ChatGPT's prompt engineering, I would write my request more specifically from the beginning.}''
\end{quote}
Therefore, proposing a specific task based on the research question is a viable solution.

\subsubsection{\textbf{Prompts Provided: Focus on Analytical Process}}
To achieve better results, it's necessary to provide ChatGPT with a more specific task description. Adding instructions in the prompt on how the AI should complete the task, such as which method to use or based on which theories, is one of the approach. For example, the prompts of P16 required ChatGPT to conduct the analysis in conjunction with the Job Demands-Resources Model~\cite{bakker2007job}. Also, we can instruct ChatGPT to read through the entire dataset, note down recurring topics, cluster these into broader themes, and present each theme with representative data excerpts and a brief summary (e.g., as an example of P9's. This process-guidance will help the AI model to generate outputs that align with our research methodology. 


\subsubsection{\textbf{Prompts Provided: Define the Format of the Inputs}}

ChatGPT performs better when analyzing formatted data compared to disorganized data~\cite{doi:10.1080/08982112.2023.2206479}, indicating that pre-cleaning of data is necessary. Cleaning the data beforehand ensures that ChatGPT is processing only the most pertinent and reliable information, thereby maximizing the value of its analytical capabilities. However, thanks to ChatGPT's capabilities in understanding context and its overall robustness~\cite{jiao2023chatgpt}, the preparation of formatted datasets can be less stringent compared to traditional data cleaning methods~\cite{chu2016data,white2012management}. The most critical aspect in preparing the dataset for ChatGPT is differentiating the information ownership, i.e., who said what. 

In addition to using a structured corpus, describing the nature and structure of the input data within the prompts is equally important. The prompt should clearly express the type of data ChatGPT will be analyzing, and we should elucidate the features of the data, such as its conversational structure, data structures entered into ChatGPT, and the data's complexity, like the length and roles. 

In our study, although not all participants used this prompt rule, it can effectively avoid potential errors such as discontinuity issues arising from multiple segmented inputs and misunderstandings of the corpus.


\subsubsection{\textbf{Prompts Provided: Define the Format of the Outputs}}
\begin{figure}[ht]
  \centering
  \includegraphics[width=0.8\linewidth]{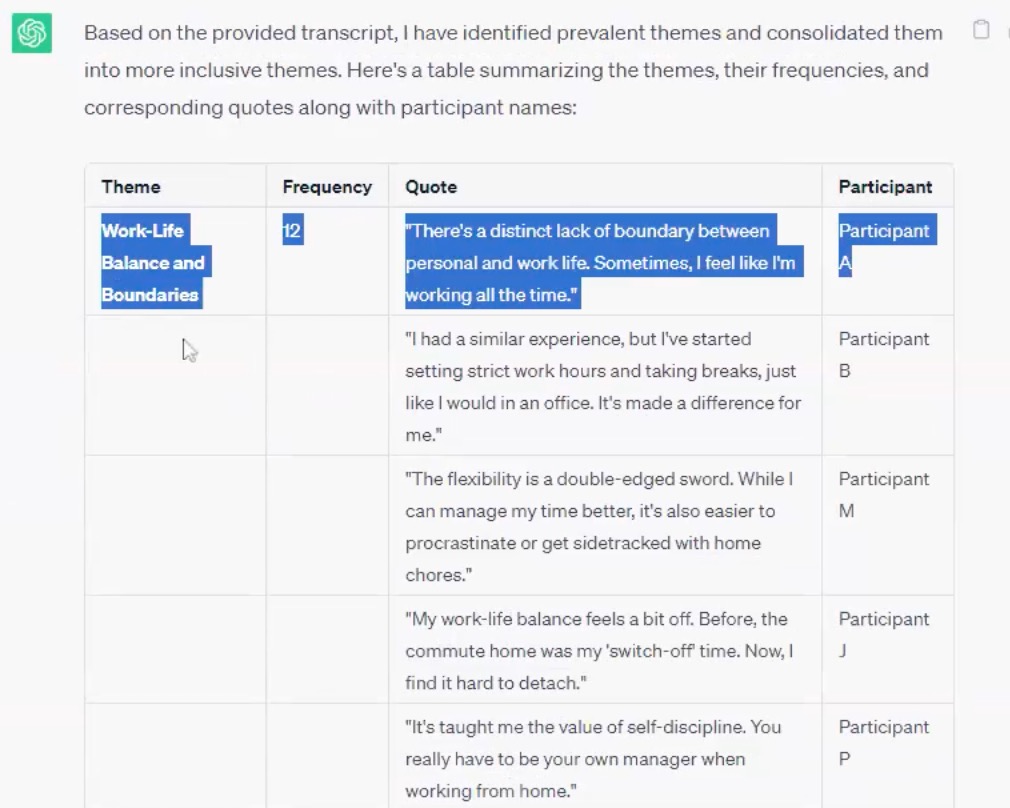}
  \caption{An example of some results: Participant (P15) added requirements regarding the format of the output results in the prompts. The original prompt text reads: ``\textit{For the output, put the result in a table. The first column is the name of the theme, the second column is its frequency, the third column includes the quotes that belong to this theme and the name of the participant who made this comment, one row per quote.}''.}
  \label{fig.outputformat}
\end{figure}

\begin{figure}[ht]
  \centering
  \includegraphics[width=0.8\linewidth]{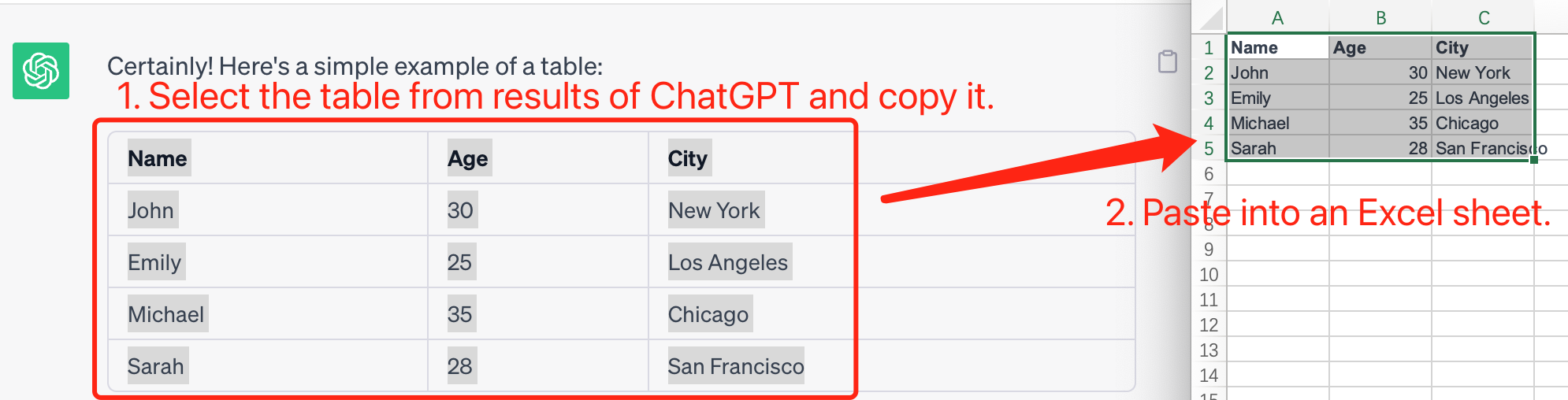}
  \caption{An example of how to transfer the table from ChatGPT to Excel.}
  \label{fig.sample_table}
\end{figure}

Apart from defining the format of input data, specifying the format of output content is equally important. Setting a standardized output format is a common strategy. Such standardized formats enhance the consistency of ChatGPT's output for the task. Notably, while not all participants explicitly defined the output format in their prompts, the results from the formal study and from a user perspective show that satisfactory outcomes have a more readable format, which is deemed beneficial for quick reading and subsequent analysis. In the formal study, some participants adopted the suggestions and requested that each topic be presented alongside relevant excerpts from the input data in a specific manner (as an example shown in Fig.~\ref{fig.outputformat}, part of the output results from P15.) and summarized the topics and their significance, which further increased the transparency of the results. Moreover, specifying the format of the output, such as a table, enables users to easily transfer it to spreadsheet software like Excel. This can be achieved by selecting and copying the table result from ChatGPT, and then pasting it into an Excel sheet, facilitating further analysis, as demonstrated in the example shown in Fig.~\ref{fig.sample_table}. 


%

\subsubsection{\textbf{Prompts Provided: Role-Playing}}
Role-playing seems to be a common and effective optimization strategy. Many previous literature mentioned the use of role-playing to enhance or surpass the performance of ChatGPT~\cite{kong2023better,eager2023prompting,liu2023jailbreaking,dong2023self,yuan2023no}. Some participants also reported that they learned about using role-playing to design prompts from past literature or tutorials. Role-playing prompts allow ChatGPT to be placed in a scenario, focusing more on completing a specific type of task. In this study, participants had ChatGPT play the role of a qualitative analysis expert to handle and analyze the corpus tasks.

However, based on the practical results, using the other strategies mentioned in this article to provide detailed descriptions for prompts can replace and surpass the advantages of letting ChatGPT role-play. Using only the role-playing strategy does not enable ChatGPT to produce consistently stable results.

\subsubsection{\textbf{Prompts Provided: Prioritization}}
Participants had certain priority requirements for the results, mainly due to (1) considerations for higher readability, and (2) the elimination of some secondary information, focusing on the analysis of the main themes in the corpus. Several participants gave positive feedback on the method of asking ChatGPT to prioritize the results. Taking P10 as an example, she mentioned: \begin{quote}``\textit{I currently feel that it's a bit too much, and I might not use so many codes.}''\end{quote} Therefore, the researcher suggested that participants add requirements for the number of codes: \begin{quote}``\textit{You can tell it the number of codes you want, for example, 'I want a codebook with less than 10 codes.'}''\end{quote} 

\begin{figure}[ht]
  \centering
  \includegraphics[width=0.8\linewidth]{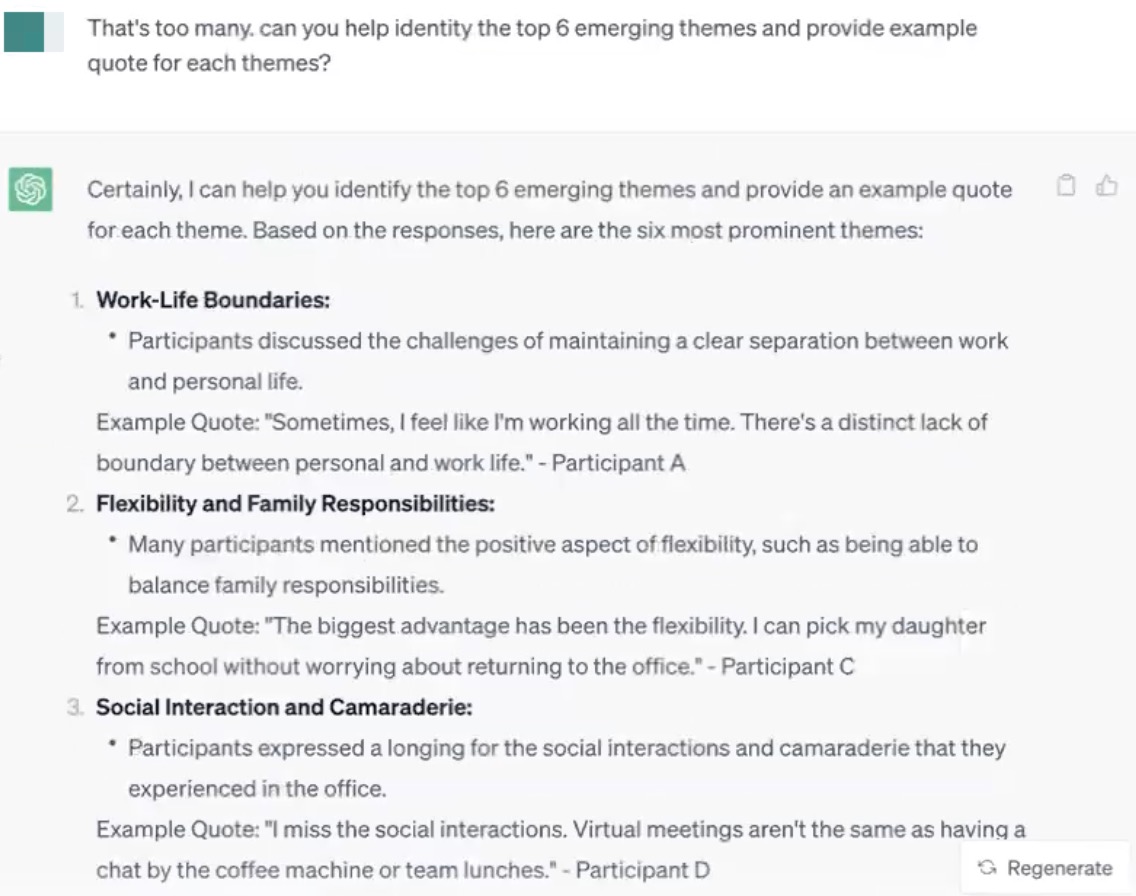}
  \caption{Examples of ChatGPT's output after adding priority requirements.}
  \label{fig.prioritization_output}
\end{figure}

Fig.~\ref{fig.prioritization_output} shows the results of ChatGPT's output after P6 used this strategy. She believes that this output result can better pinpoint key codes. At this stage, not only was the practical effect positive, but it also revealed some users' habits and personalized needs when handling tasks. We hope that the proposed framework serves as a reminder for novice users interacting with ChatGPT, helping them recall the traditional analysis process and apply ChatGPT's capabilities to enhance their analytical tasks.


\subsubsection{\textbf{Prompts Provided: Clarification, Transparency and Traceability}}
The lack of transparency is a significant challenge for generative AI~\cite{DWIVEDI2023102642}. Due to the black box, transparency problems have long been criticized by researchers and users~\cite{8466590}. The interviews in the formal study confirmed that the lack of transparency is one of the main reasons for users to be cautious about using ChatGPT for qualitative data analysis. In fact, all participants had concerns about the transparency of the content generated by ChatGPT to varying degrees. Although the black box is not entirely untrustworthy~\cite{castelvecchi2016can}, the results of formal study show that all participants demanded further clarification from ChatGPT to enhance the transparency of the results. Adding explanatory information, traceable sources, and standardized output formats can significantly increase users' trust in ChatGPT results~\cite{BARREDOARRIETA202082}. We've previously mentioned standardized output formats, and another effective method to increase transparency is to provide traceable information sources for output results~\cite{franzoni2023black}. 

When using ChatGPT for qualitative analysis of a corpus, demanding better interpretability and higher transparency may require just a single sentence, as done by P13. However, combining this prompt with other strategies can achieve better results, achieving better readability as shown in Fig.~\ref{fig.outputformat} (P15) or Fig.~\ref{fig.prioritization_output} (P6).

In this formal study, we found that asking ChatGPT to analyze each line of data independently, rather than conducting an overall analysis of the input data, is a more effective strategy. Although overall analysis can still produce usable or insightful insights, through comparison, independently analyzing each response plays a more significant role in subsequent in-depth studies and may lead to more discoveries. This doesn't mean that analysis should be conducted without considering context. A comprehensive analysis can be achieved by combining other prompt strategies, such as considering priority.

\subsubsection{\textbf{Prompts Provided: Acknowledgment of Expertise}}
\begin{figure}[ht]
  \centering
  \includegraphics[width=0.9\linewidth]{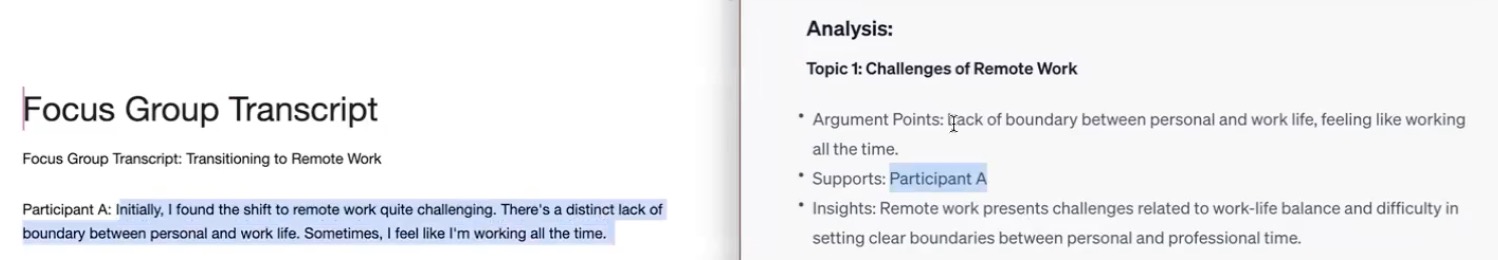}
  \caption{Examples of ChatGPT's output from P8's prompts.}
  \label{fig.thanks2chatgpt}
\end{figure}

Incentive measures can ensure that the system consistently provides the same output for the same input~\cite{kilhoffer2023ai}. In practice, participants indeed provided positive feedback, such as the results obtained by P8 (see Fig.~\ref{fig.thanks2chatgpt}), and their evaluation and interpretation of the results: \begin{quote}``\textit{(On the evaluation of the output) So smart! It still needs to be praised. (Explanation) If you compliment it, it might be a bit smarter. Because ChatGPT's memory is limited, you should praise it a bit. Otherwise, it might get `dumber' the more you use it, so you need to give it a compliment.} (P8)''\end{quote} However, researchers believe that this depends more on other more specific strategies (such as setting a standardized output format, requiring original information, etc.). In the practice of P17, mere incentives did not noticeably improve the quality of the generated results, but adding other strategies to the prompts clearly enhanced the readability of the results and user trust. However, P17 gave us a new insight: using polite and encouraging words can improve AI's ``learning environment.'' P17 mentioned: 
\begin{quote}
    ``\textit{Because I think whatever I say to the AI, that's what the AI will eventually become. I want a polite and kind AI. I hope to provide some good data references for training the AI.}''
\end{quote}


\subsubsection{\textbf{Iteration of Prompts}}

Participants recognized the ability of interacting with LLMs like ChatGPT via natural language, as noted by P7, \textit{``I feel the advantage of using ChatGPT lies in this conversational aspect, where depending on your input, you might get slightly different outcomes every time.''} While detailed procedures were helpful for better utilization of ChatGPT, participants did not want to confine their interactions to specific commands. This approach not only made the framework more versatile, but also allowed users to incorporate their own critical and creative thinking into the interaction process~\cite{paul2004critical}. Specifically, users could critically evaluate and judge the results provided by ChatGPT, while also infusing more creativity into the design of the prompts.

LLMs-based applications like ChatGPT power a new generation of interactive AI applications, introducing a unique user interaction paradigm. They allow users to refine outputs through iterative prompt design, addressing various needs, such as offering explanations, elaborating queries, or facilitating knowledge transfer. This capability to reconfigure prompts and improve outcomes is undeniably one of ChatGPT's most potent attributes. In our study involving qualitative analysis tasks using ChatGPT, participants consistently drew inspiration from the iterative prompt design. Furthermore, the strategies we've detailed above benefited immensely from this iterative approach.

Participants recognized the ability to interact with LLMs like ChatGPT via natural language, as noted by P7:

\begin{quote}
\textit{"I feel the advantage of using ChatGPT lies in this conversational aspect, where depending on your input, you might get slightly different outcomes every time."}
\end{quote}

While detailed procedures were helpful for better utilization of ChatGPT, participants did not want to confine their interactions to specific commands. This approach not only made the framework more versatile but also allowed users to incorporate their own critical and creative thinking into the interaction process~\cite{paul2004critical}. Specifically, users could critically evaluate and judge the results provided by ChatGPT while also infusing more creativity into the design of the prompts.

The iterative nature of prompt design and the user's critical assessment of the outputs are essential elements when using LLMs for qualitative data analysis. Our framework incorporates these aspects, encouraging users to engage in multiple iterations of prompt refinement and critically evaluate the results at each stage. This iterative process allows users to fine-tune the prompts, ensuring that the LLM's outputs align with their research objectives and provide meaningful insights.

However, multiple iterations might diminish ChatGPT's performance in processing the original task, context, or specific tasks, as P8 pointed out:
\begin{quote}
\textit{"[With multiple iterations,] it might lose focus on the original task or context."}
\end{quote}
This indicates that if we are going to engage in multiple iterations, we need to ensure that the LLM still has a clear understanding of its objectives after each iteration. Therefore, we recommend that during the iterative process, the other parts mentioned in the framework should be simultaneously or repeatedly emphasized to ensure consistency in the output results.

\subsubsection{\textbf{Robustness}}\label{subsec:robustness}
\begin{figure}[ht]
	\begin{minipage}[htbp]{0.48\linewidth}
		\centering
		\includegraphics[width=1\linewidth]{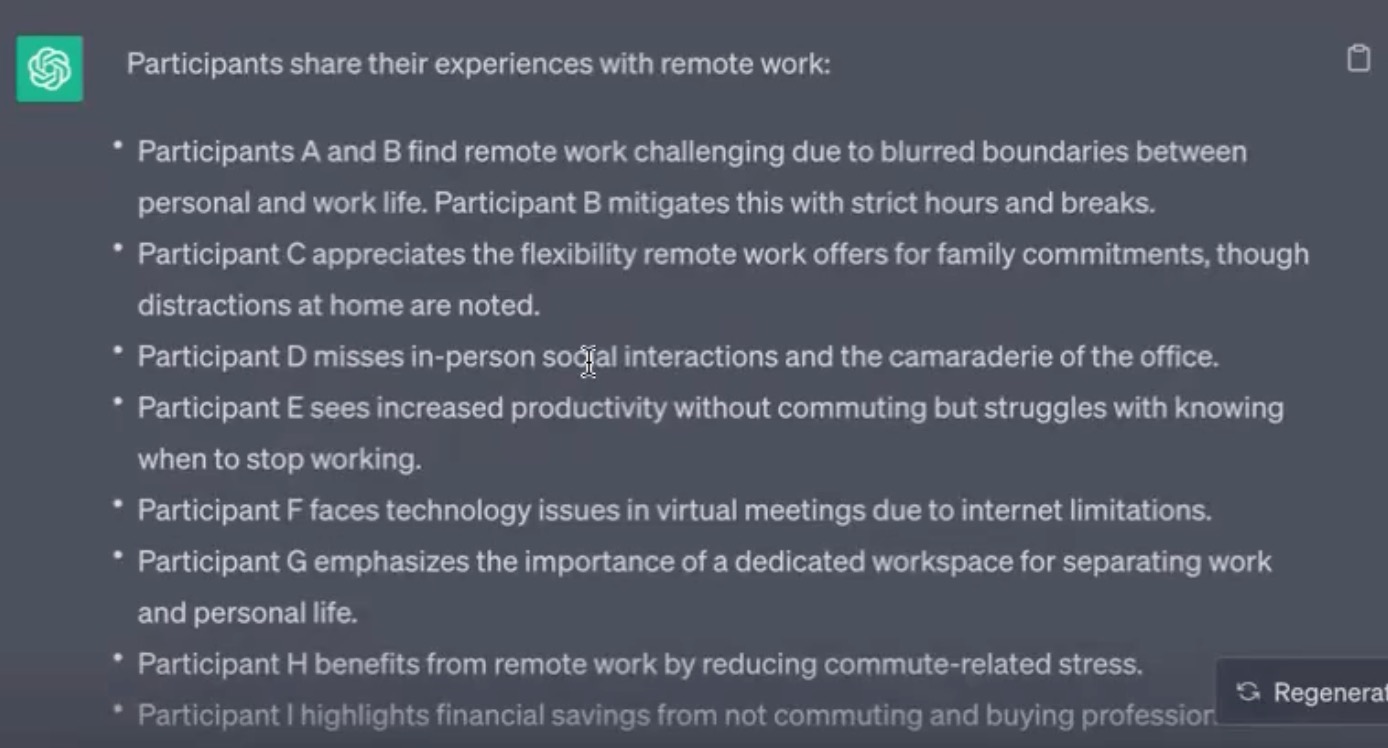}
	\end{minipage}
	\begin{minipage}[htbp]{0.48\linewidth}
		\centering
		\includegraphics[width=1\linewidth]{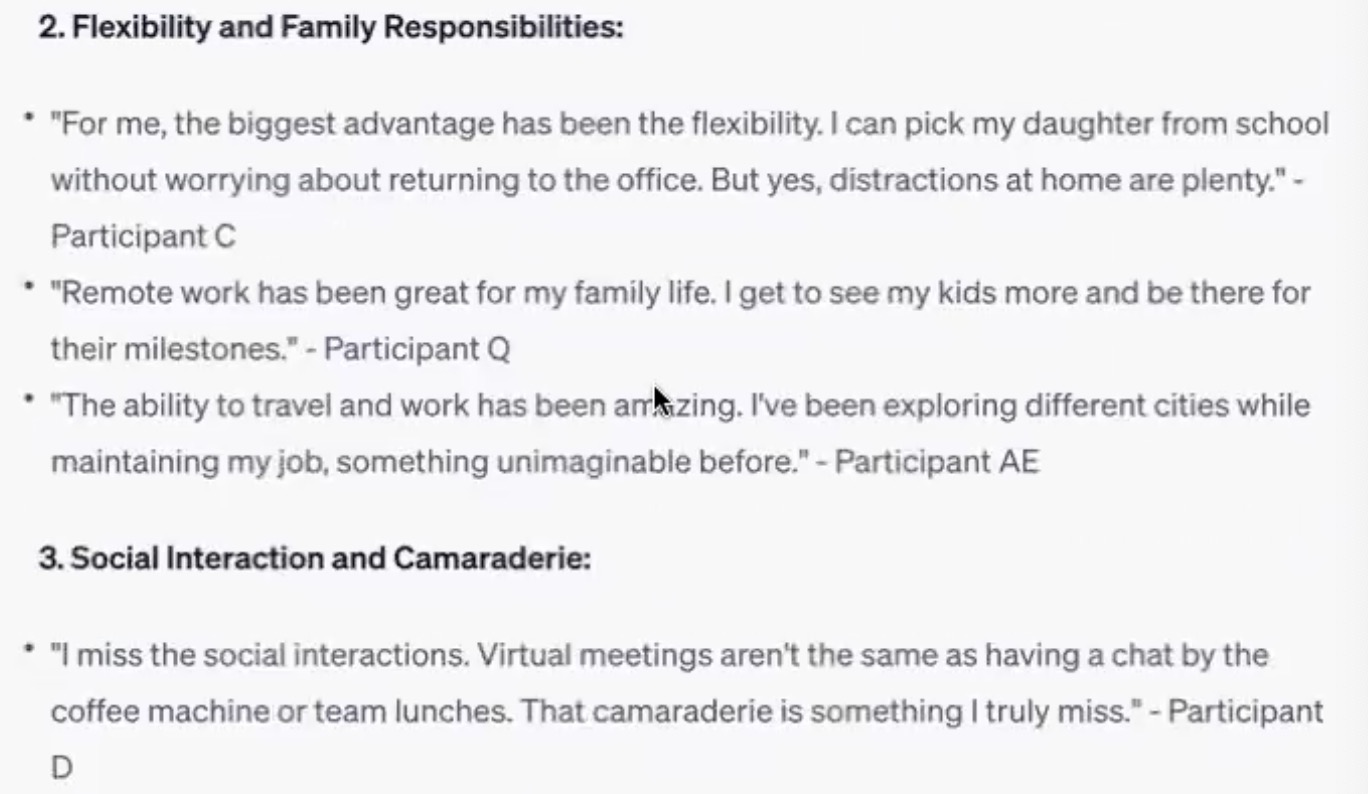}
	\end{minipage}
	\caption{Some of the results obtained by participants using ChatGPT. On the left is the result for P5, and on the right is the result for P6.}
	\label{fig.robustnessoutput}
\end{figure}

As mentioned earlier, some participants did not input certain features in the prompts, yet the results generated by ChatGPT met or exceeded their expectations, such as the output format being readable and the addition of the source of information (as shown in Fig.~\ref{fig.robustnessoutput}, part of the output results from P5 and P6.). This may be due to (1) the relevance of features in the prompts, and (2) the robustness of ChatGPT to the task~\cite{roumeliotis2023chatgpt}.


\textbf{The Relevance of Features:}
We noticed that prompts carefully designed by numerous participants frequently resulted in similar outcomes. For instance, P6 and P7 directed ChatGPT to role-play but did not specify a certain output format. In stark contrast, P14 and P15 delineated an output format but steered clear of the role-playing technique. This observation hints at a potential correlation between the strategies of role-playing and output-format specification, suggesting they might be interchangeable in certain scenarios.

\textbf{Interacting using Natural Language:} The ChatGPT, based on LLMs, allows users to interact using natural language, which is one of the most significant advantages of this class of AI applications. Since ChatGPT can understand natural language, the prompts for interacting with it do not need to be fixed instructions or code.
 \begin{quote}
 ``\textit{It's like having a conversation with someone}’‘ (P13)
\end{quote}
In this context, a framework for designing prompts for qualitative analysis tasks appears to be more valuable than a fixed, single-purpose command. On one hand, users no longer need to remember specific codes or formulas; instead, they interact with ChatGPT in a more flexible manner based on their understanding of the framework. This greatly reduces the learning cost for learners. On the other hand, this framework provides a minimum level of satisfactory performance output, offering users a high degree of freedom in its use. This freedom is significant for creative purposes, enabling users to create better prompts in the future, explore and enhance ChatGPT's capabilities in qualitative analysis or other fields, and collaborate more effectively with ChatGPT.

\subsection{Notes on ChatGPT with Different Versions}
In the course of our research, we opted to use ChatGPT based on the GPT-3.5 model rather than the 4.0 version, primarily for reasons of accessibility: 1. During the period of formal study, the GPT-4.0 version of ChatGPT had a maximum access limit per four hours; 2. The 4.0 version of ChatGPT is a subscription service, and while we encourage subscriptions, it should not be a mandatory choice for this research. Furthermore, although OpenAI claims that ChatGPT 4.0 has improved performance, prior research found that prompting GPT-4 to walk through the ingredient list item-by-item proved challenging enough~\cite{10.1145/3563657.3596138}. Moreover, based on our tests, the quality of the results did not demonstrate a significant improvement in performance for GPT-4.0 over the GPT-3.5 version of ChatGPT. In other words, there was little difference in the outputs generated for the tasks in this study.

\section{User's Attitude on ChatGPT's Qualitative Analysis Assistance: From No to Yes}

Participants initially expressed significant concerns about ChatGPT's transparency (interpretability and verifiability), performance (consistency and accuracy), the difficulty of designing prompts, and the cost of reviewing results. However, after applying the framework we suggested, participants reported a substantial increase in their confidence when interacting with ChatGPT and in its ability to process qualitative data effectively.

\begin{quote}
``\textit{After designing the prompts step by step, I think both the format and content have been improved, such as the source of the data and the ability to integrate it into a table format... If I were to use ChatGPT for batch coding, I would use such a template [framework].}'' (P15)
\end{quote}
This quote illustrates how the structured prompt design framework enhanced the transparency and usability of ChatGPT's outputs for qualitative analysis tasks.

\begin{quote}
``\textit{From my experience today, I feel its [ChatGPT's] summarization ability is quite strong. Also, I hadn't thought before that qualitative data could be analyzed with it.}'' (P7)
\end{quote}
This suggests that the framework not only improved the quality of ChatGPT's responses but also expanded participants' understanding of its potential applications in qualitative research.

The primary reason for this shift in attitude was that the framework-guided design of prompts effectively enhanced the transparency and credibility of ChatGPT's outputs. By providing a structured approach to prompt design, the framework enabled participants to elicit more interpretable and verifiable responses from the AI model, thus increasing their trust in its capabilities.

These findings demonstrate the transformative impact of a human-centric prompt design framework on qualitative researchers' attitudes towards using ChatGPT in their work. The framework's ability to address key concerns about transparency, performance, and usability suggests that it could play a crucial role in facilitating the adoption of AI tools in qualitative research practices.

\subsection{Attitude Shift Due to Increased Transparency and Trust}

After applying the prompt design framework, participants reported a substantial increase in their confidence when interacting with ChatGPT and in its ability to process qualitative data effectively. The framework's emphasis on transparency, particularly in linking ChatGPT's responses to specific data sources, enhanced participants' trust in the tool.

\begin{quote}
``\textit{In terms of preliminary screening and categorization, it has already saved a lot of time, which I think is a very okay process. Another great thing is that I know its data source... I can also trace back to see who made this point and what it originally meant.}'' (P8)
\end{quote}

Despite this improvement, participants still recognized the importance of human oversight in the process.

\begin{quote}
``\textit{I think the results from ChatGPT are very good. And when presenting these [analysis results], it even connects them with the user, which I find very intuitive. This [format] will increase my trust in ChatGPT. Still, I won't directly state my research results based solely on its findings. I won't do that; I would definitely double-check. [Overall,] I think ChatGPT is exceptionally useful.}'' (P5)

\end{quote}

This quote illustrates that while the framework significantly improved ChatGPT's transparency and credibility, researchers still valued the role of human judgment in validating the AI's outputs.

These findings suggest that the prompt design framework's ability to increase transparency in ChatGPT's responses was a key factor in shifting participants' attitudes from skepticism to acceptance. By providing a structured approach to elicit more interpretable and verifiable outputs, the framework fostered greater trust in the AI tool's capabilities while still acknowledging the importance of human oversight in the qualitative analysis process.

\subsection{Expanded Understanding of ChatGPT's Potential}

The framework not only improved the quality of ChatGPT's responses but also expanded participants' understanding of its potential applications in qualitative research.

\begin{quote}
``\textit{From my experience today, I feel its [ChatGPT's] summarization ability is quite strong. Also, I hadn't thought before that qualitative data could be analyzed with it.}'' (P7)
\end{quote}

By designing prompts, participants further understood the capabilities of ChatGPT and changed their attitude towards using ChatGPT for qualitative analysis tasks. It can be seen that the concerns of qualitative researchers about using ChatGPT to process data can be addressed through well-designed prompts. Based on the shift in participants' attitudes before and after applying the prompt design framework, it is evident that the prompt framework we designed (presented in the section~\ref{sec:promptdesignprocess}) is beneficial for utilizing LLMs to assist qualitative analysis work.

\section{Discussion}
Our research results show that LLMs-based applications (such as ChatGPT) have the potential to conduct qualitative analysis on corpora through well-designed prompts, addressing concerns of human analysts. In the following sections, we will expand on the discussions by Jiang et al.~\cite{10.1145/3449168} and Feuston and Brubaker~\cite{10.1145/3479856} about the collaboration between humans and AI in qualitative research, especially for junior researchers. Our inspiration comes from the views of the participants and is drawn from the processes and methods of qualitative research ~\cite{jensen2013handbook}.

In addition, we will discuss the ethical considerations of using ChatGPT from the user's perspective and its impact on the conceptual transition process (from rejection to acceptance). We draw inspiration from Idhe's book ``Instrumental Realism''~\cite{ihde1991instrumental}, concepts from Husserl~\cite{husserl2012ideas}, and ideas from Foucault~\cite{foucault1988technologies}. Combining practice, tools, and phenomenology, we discuss how new AI tools like ChatGPT can influence user attitudes by shifting paradigms.





\subsection{Overcoming the Challenges in Prompt Design}

Our findings highlight the difficulties researchers face when designing prompts for ChatGPT to support qualitative analysis tasks, particularly for junior researchers who lack expertise and experience in prompt design \cite{10.1145/3544548.3581388,fiannaca2023Programming}. To address these challenges, we propose a framework that incorporates strategies identified in our findings, such as providing background information, focusing on methodology and analytical process, defining data formats, role-playing, prioritization, ensuring transparency, and acknowledging expertise. These strategies align with existing research on prompt engineering techniques \cite{NEURIPS2020_1457c0d6, zhao2021Calibrate, wei2022ChainofThought, gao2023Prompt}.

The framework aims to make the prompt design process more structured and transparent, enabling researchers to elicit more interpretable and verifiable responses from ChatGPT. This approach addresses concerns regarding the lack of transparency in ChatGPT's outputs \cite{10.1145/3641289} and aligns with research on explainable AI (XAI) \cite{10.1145/3387166}. Moreover, the framework improves junior researchers' understanding of ChatGPT's capabilities in qualitative analysis by providing guidance on effectively communicating context, methodology, and data formats to the AI \cite{10.1145/3449168, katz2023Exploringa, siiman2023Opportunities}.

The framework's approach also considers the specific challenges associated with thematic analysis \cite{castleberry2018thematic, guest2013collecting, morgan2022understanding, 10.1145/3544548.3580766, braun2019reflecting}. By leveraging ChatGPT's capabilities through effective prompt design, researchers can potentially mitigate these challenges and enhance the efficiency of the thematic analysis process. However, it is crucial to recognize the importance of human oversight and validation in the qualitative analysis process \cite{10.1145/3479856, 10.1145/3449168, christou2023Iow, 10.1145/3526113.3545681, 10.1145/3544548.3581352}.

In summary, our proposed framework addresses the challenges in prompt design by providing a structured approach that incorporates strategies to improve transparency, understanding, and effectiveness of ChatGPT-assisted qualitative analysis. Future research is needed to further refine and validate the framework, address potential challenges, and explore the optimal integration of AI-assisted tools into the qualitative research process.

\subsection{The Robustness of ChatGPT}
In addition to overcoming challenges of prompting, our research also uncovered the ``robustness'' of ChatGPT during interactions, which is the system's ability to handle unforeseen inputs or conditions without failing or producing erroneous results \cite{159342}. \citet{dietterich2019robust} argued that the more powerful technology becomes, the more important is the robustness. Based on participants' feedback, ChatGPT has demonstrated robustness in at least two aspects during user interactions : (1) Ability to Understand Natural Language: Users can communicate with ChatGPT using everyday language, without the need for any specific commands or programming code. (2) Error Tolerance: ChatGPT is highly resistant to user input errors, such as spelling or grammatical mistakes~\cite{roumeliotis2023chatgpt}. It can effectively understand input with errors and provide meaningful responses. (3) Efficiency and effectiveness: ChatGPT responds to prompts and processes data quickly. Participants typically reach the desired results within limited number of steps regardless of their prompt strategies. (4) Transparency: Despite the complexity of LLMs, ChatGPT is able to provide explanations and details about its decision-making process. With appropriate prompts delineating the output format, ChatGPT can make the data analysis transparent to users. 

Still, the robustness of ChatGPT's could be further enhanced through strategically designed prompt interactions. Specifically, well-designed prompts could improve the following often-criticized areas: (1) The readability and consistency of output results: By clearly defining the output format, we can ensure that the responses are in a highly consistent and readable format, which facilitates the storage of results. (2) Transparency and Interpretability: By requiring the source or original text of the corresponding raw data, the results from ChatGPT become verifiable, which helps users to track the analysis process and improve the reproducibility of doing qualitative analysis with ChatGPT.

\subsection{Methodology for Incorporating ChatGPT into Thematic Analysis: A Framework of Prompts Design}\label{subsec:mergingprompts}

In our study, we integrated established prompt design methodologies~\cite{10.1145/3563657.3596138} to enhance ChatGPT's efficacy in processing structured data. We addressed the challenge of command failures, as highlighted by Zamfirescu-Pereira et al.~\cite{10.1145/3563657.3596138}, not by simply appending multiple prompts, but by strategically integrating them and simplifying the interaction process.

Post-experiment, we distilled the prompt techniques agreed upon by researchers and participants into a streamlined framework, as depicted in Fig~\ref{fig.teaserfigure}. There are four core parts in the framework: 1) Description of the Task's Background. This part enables users to set the context and provide information that helps ChatGPT to understand the nature and structure of the input data. 2) Description of the Task. This part gives further instructions about the task that ChatGPT needs to execute with the input data. 3) Description of how the task will be processed. This part serves as a methodological guidance to ensure that ChatGPT will complete the task following a specific procedure. 4) Description of the Expected Output Content. Lastly, this framework will instruct ChatGPT to organize its responses in a reproducible and easily transferable format. Following this framework, all requirements of qualitative analysis, individual feedback processing, and theme categorization can be combined into a singular, comprehensive prompt. This prompt enables users to exploit ChatGPT's context retention capability effectively, thereby avoiding repetitive inquiries. In addition, this is a flexible framework that allows integration of customized instructions such as Role-Playing, and pre-defined codebook.

By employing this framework  rather than specific instructions or code, users can craft more efficient ChatGPT prompts for qualitative analysis. The robustness of ChatGPT permits users to leverage the framework's recommendations through natural language, optimizing the output quality.

\subsection{Improving Junior Researchers' Understanding of ChatGPT's Capabilities in Qualitative Analysis}

Our findings and proposed framework highlight the potential for ChatGPT to support junior researchers in understanding and conducting qualitative analysis. By providing a structured approach to prompt design that emphasizes context, methodology, data organization, and transparency, the framework helps junior researchers develop a deeper understanding of qualitative research methods and best practices, addressing challenges such as subjectivity in identifying themes \cite{morgan2022understanding}, varying interpretations \cite{10.1145/3544548.3580766}, and the need for reflexivity and transparency \cite{braun2019reflecting}.

For junior researchers lacking formal training or experience in qualitative research methods, the framework serves as a valuable learning tool, guiding them through the process of designing effective prompts for ChatGPT and breaking down the complexities of thematic analysis into manageable steps. By leveraging ChatGPT's capabilities, junior researchers gain hands-on experience in applying qualitative analysis methods to real-world data, developing a more intuitive understanding of identifying themes, patterns, and insights within qualitative data.

The framework's emphasis on transparency and explainability helps junior researchers develop a more critical and reflective approach to qualitative analysis, encouraging them to interrogate assumptions and biases embedded in the AI's outputs. Furthermore, the efficiency gains afforded by ChatGPT and the framework lower barriers to entry for junior researchers, democratizing access to qualitative research methods and encouraging more junior researchers to incorporate these methods into their work, particularly those outside the field of qualitative methods.

As more junior researchers across diverse disciplines leverage AI-assisted tools like ChatGPT in their qualitative research, we may see a proliferation of new applications and innovations in qualitative analysis. The framework's flexibility and adaptability to different research contexts can facilitate cross-pollination and interdisciplinary collaboration, as researchers from various backgrounds learn from and build upon each other's experiences and insights.

\subsection{The Evolving Landscape of Ethical Considerations}
The ethical quandaries presented by the confluence of human intelligence and AI have long been a focus of both scholarly deliberation and real-world application~\cite{10.1145/3491102.3517732,10.1145/3491102.3502103}. Our study revealed a significant transformation in participants' attitudes towards ChatGPT's use in qualitative research. Initially skeptical, they progressively moved towards endorsing ChatGPT, which further shed light on the evolving landscape of these ethical considerations, highlighting the shifts in human trust and acceptance in response to the transparency and efficacy of AI tools. 

 This shift is not solely attributable to the technical prowess of ChatGPT but also highlights the role of our framework in addressing key ethical concerns, particularly those related to transparency and accountability. This observation aligns with the concept that the real value of technology is realized through user experience and functionality comparison~\cite{hassenzahl2010experience}.

Describing the relationship between the human analyst and ChatGPT, one can envisage an intricate, symbiotic interaction~\cite{10.1007/978-3-319-13500-7_1}. In this context, ChatGPT is not just an auxiliary tool; it becomes an extension of the human cognitive process~\cite{JARRAHI2018577}, enhancing analytical capabilities while simultaneously reshaping one's epistemological outlook.

Examining this relationship through Husserl's phenomenological lens deepens our understanding~\cite{husserl2012ideas}. Interaction with technology, in this view, is not just transactional but a profound, lived experience. Participants in our study did not only simply use ChatGPT, but they also engaged with it, evolving their understanding and establishing trust in its analytical capabilities. Foucault's notion of ``technologies of the self'' further enriches this discussion~\cite{foucault1988technologies}. Seen through this lens, ChatGPT serves not only as an analytical tool but also as a catalyst for personal and professional transformation. It encourages scholars to reconsider their methodologies and the very limits of knowledge~\cite{10113601}. Learning knowledge from the interaction process and reflecting on it is of great significance, especially for early-career individuals and student groups.

Additionally, the integration of human intelligence with AI, particularly in the context of qualitative analysis, brings forth profound ethical considerations that demand meticulous attention. Informed by the shift in researchers' attitudes as revealed in our findings, it becomes evident that the ethical dimensions of this integration are multifaceted. Nissenbaum's emphasis on context in shaping our ethical expectations from technology is highly pertinent here~\cite{nissenbaum2010privacy}. The shift from skepticism to acceptance among researchers in our study underscores a growing reliance on tools like ChatGPT. While this reliance speaks to the efficiency and potential of AI, it simultaneously raises concerns about the unintentional reinforcement of biases and the potential narrowing of the scope of qualitative inquiry. The construction and contemplation of ethical expectations for AI is not only timely for novice researchers, but also conducive to stimulating more in-depth reflections.

The ethical implications of this scenario are twofold. First, there is a risk that AI systems, including ChatGPT, might inadvertently perpetuate existing biases in their training data~\cite{motoki2023more}. This highlights the crucial importance of training data, especially in qualitative research, where a nuanced understanding and interpretation of data are key. Biased or toxic training data could lead to the development of an AI that is ``unfriendly'' and ``unethical''. As the influence of AI in everyday life increases, and human-AI interactions become more frequent, such a ``toxic'' AI could harm human interests in various ways, including the quality of interaction experiences, emotional responses, and moral perspectives. Second, the growing trust in AI’s capabilities may lead to a diminished emphasis on the critical, reflective role traditionally played by human analysts in qualitative research. This shift might result in less depth in analysis, as AI tools may not fully replicate the complex cognitive processes inherent in human analysis~\cite{koivisto2023best}.

Therefore, our findings not only highlight the growing acceptance and utility of AI in qualitative research but also call for a balanced approach. This approach should ensure that while AI tools like ChatGPT augment and assist in research, they do not overshadow the indispensable human elements of intuition, skepticism, and ethical judgement that are crucial in qualitative analysis.

\section{Limitations and Future Work}
In charting new territory by employing LLM-based applications, such as ChatGPT, to support qualitative analysis, we recognize the inherent constraints of our exploratory approach. Acknowledging these limitations not only brings transparency to our current findings but also paves the way for refining subsequent research endeavors.

\textbf{L1. Corpus Scope and Diversity:} The specificity of our selected corpus may introduce limitations. Designed with the intent of facilitating the study, this corpus might not capture the expansive nuances and varieties inherent in broader qualitative data. As such, we posit that incorporating diverse qualitative corpora in future research could significantly amplify the generalizability and applicability of our findings.

\textbf{L2. Long-term Implications of ChatGPT Interactions:} Our study offers insights based predominantly on short-term experiments, with durations just exceeding an hour. The evolving demands of qualitative analysts—driven by advancements in research methodologies or prolonged, iterative interactions with tools like ChatGPT—were outside the purview of this study. It's plausible that with consistent and deepened interactions, ChatGPT and similar models might undergo adaptive learning, culminating in new version iterations. Thus, our current insights and proposed frameworks should be viewed as foundational, given the embryonic nature of this domain. There's an implicit understanding that our contributions may need revisiting and refining with the evolution of ChatGPT or the emergence of newer AI tools.

\textbf{L3. Performance Across Qualitative Analysis Tasks:} This research predominantly confirms ChatGPT's great power in the coding facet of qualitative data analysis. However, its efficacy across the broader spectrum of tasks inherent to qualitative research remains largely unprobed. Delving deeper into these uncharted territories in subsequent studies will offer a more holistic understanding of LLMs in qualitative research.

In essence, our foray into the interplay between LLMs like ChatGPT and qualitative analysis is just the tip of the iceberg. While we've laid down some initial markers, there's a vast expanse yet to be explored, promising exciting prospects for future interdisciplinary research. Several promising directions for future work include:

\textbf{Fw1. Expanding ChatGPT's Capabilities to the Humanities:} We draw inspiration from the work of Marcoci et al.~\cite{marcoci2023big}, recognizing that often, when discussing technological methodologies and applications, the voices of researchers from non-STEM disciplines are overlooked. There's both a need and a potent opportunity to extend AI technologies and applications, like ChatGPT, to the humanities. Doing so would not only enhance interdisciplinary research but also ensure that technology is inclusive and resonates with broader academic communities.

\textbf{Fw2. Ethical Considerations of LLM-based Applications:} Our findings have prompted us to further delve into the ethical implications and concerns surrounding applications based on LLMs. While we explained the alleviation of ethical concerns through the lens of a paradigm shift, delving deeper into these ethical dimensions is crucial for ensuring the responsible advancement of AI. It also paves the way for advancing societal understanding and fostering a more informed discourse on the integration of AI in various domains.

\textbf{Fw3. Personalized and Integrated LLM Toolkit:} One of the desires expressed by qualitative analysts (P7, P9, P15, P16) is the creation of an integrated LLMs toolkit tailored specifically for qualitative analysis. We have taken this recommendation into account for our future work trajectory. Furthermore, insights from one of our study participants (P4) regarding expectations for future iterations of ChatGPT have been enlightening. While our results show that ChatGPT can effectively analyze qualitative data, it's crucial to acknowledge the vast diversity in user backgrounds. Coupled with the complexities tied to each cultural context, it underscores the importance of AI technologies being attuned to regional and cultural nuances. Such cultural cognizance not only ensures the pertinence of the produced content but also augments user trust and engagement. Moving forward, there is potential to develop a version of ChatGPT that amalgamates personalized knowledge bases, thereby catering to the distinct needs of varied global communities. In addition, future updates of GPT might affect the performance of processing such tasks. Based on the latest user feedback after the update to the GPT-4.0-Turbo version, we have to admit that some updates might negatively optimize the application performance. Therefore, establishing a fixed-purpose LLMs application significantly reduces the costs associated with changing usage methods and decreases the uncertainty of the results.

\textbf{Fw4. Similar Strategies, Broader Applications: Extended Applications and Capabilities of ChatGPT:} ChatGPT's integration within our framework has demonstrated proficiency in qualitative analysis, but its potential extends into various other domains, e.g., quantitative analysis, programming, and creative writing, etc 
\textbf{Fw5. Exploring the Role of ChatGPT: Tool or Co-researcher?}
One of future work should investigate the potential roles of ChatGPT in qualitative analysis as a tool or a co-researcher. As a tool, ChatGPT can streamline coding processes and offer new insights, but may lead to over-reliance on AI. As a co-researcher, ChatGPT could actively contribute to thematic analysis, but raises questions about AI "understanding" and ethical considerations. Future studies should examine how these roles impact the quality and efficiency of qualitative analysis, develop prompt design frameworks and collaboration strategies, and address the ethical implications of human-AI collaboration. Recognizing ChatGPT's strengths and limitations and employing it to amplify the quality and depth of qualitative research will be crucial for advancing AI-assisted qualitative analysis.

\section{Conclusion} 
This study first identified the risks and challenges of ChatGPT in qualitative analysis through a pilot study. It then explored the attitudes of a group of qualitative analyst towards the application of ChatGPT in qualitative research through interviews and experiments, and in collaboration with this study group, developed a well-received framework for prompts design. Our research explored the powerful capabilities of AI in qualitative analysis using ChatGPT, which could potentially significantly reduce the labor-intensive tasks and coding costs in qualitative analysis in the future. Our research findings indicate that enhancing transparency, providing guidance on prompts, and strengthening users' understanding of LLM capabilities can significantly improve user interaction with ChatGPT and reverse negative attitudes towards using such applications in research. In the discussion, we focus on the challenges, potential, and impact on novice researchers associated with the application of ChatGPT, based on our findings. Furthermore, we delve into the ethical considerations brought about by advanced AI, especially in the context of qualitative analysis, starting from the shift in user attitudes observed during the research process. Lastly, we proposed several pressing future works to further expand, delve into, and understand LLMs, providing insights. We hope that this study can help users better apply new technologies to enhance efficiency.
\begin{acks}

\end{acks}

\bibliographystyle{ACM-Reference-Format}
\bibliography{sample-base}


\begin{thebibliography}{109}


\ifx \showCODEN    \undefined \def \showCODEN     #1{\unskip}     \fi
\ifx \showDOI      \undefined \def \showDOI       #1{#1}\fi
\ifx \showISBNx    \undefined \def \showISBNx     #1{\unskip}     \fi
\ifx \showISBNxiii \undefined \def \showISBNxiii  #1{\unskip}     \fi
\ifx \showISSN     \undefined \def \showISSN      #1{\unskip}     \fi
\ifx \showLCCN     \undefined \def \showLCCN      #1{\unskip}     \fi
\ifx \shownote     \undefined \def \shownote      #1{#1}          \fi
\ifx \showarticletitle \undefined \def \showarticletitle #1{#1}   \fi
\ifx \showURL      \undefined \def \showURL       {\relax}        \fi
\providecommand\bibfield[2]{#2}
\providecommand\bibinfo[2]{#2}
\providecommand\natexlab[1]{#1}
\providecommand\showeprint[2][]{arXiv:#2}

\bibitem[159(1990)]%
        {159342}
 \bibinfo{year}{1990}\natexlab{}.
\newblock \showarticletitle{IEEE Standard Glossary of Software Engineering Terminology}.
\newblock \bibinfo{journal}{\emph{IEEE Std 610.12-1990}} (\bibinfo{year}{1990}), \bibinfo{pages}{1--84}.
\newblock
\urldef\tempurl%
\url{https://doi.org/10.1109/IEEESTD.1990.101064}
\showDOI{\tempurl}


\bibitem[Adadi and Berrada(2018)]%
        {8466590}
\bibfield{author}{\bibinfo{person}{Amina Adadi} {and} \bibinfo{person}{Mohammed Berrada}.} \bibinfo{year}{2018}\natexlab{}.
\newblock \showarticletitle{Peeking Inside the Black-Box: A Survey on Explainable Artificial Intelligence (XAI)}.
\newblock \bibinfo{journal}{\emph{IEEE Access}}  \bibinfo{volume}{6} (\bibinfo{year}{2018}), \bibinfo{pages}{52138--52160}.
\newblock
\urldef\tempurl%
\url{https://doi.org/10.1109/ACCESS.2018.2870052}
\showDOI{\tempurl}


\bibitem[Alkaissi and McFarlane(2023)]%
        {alkaissi2023artificial}
\bibfield{author}{\bibinfo{person}{Hussam Alkaissi} {and} \bibinfo{person}{Samy~I McFarlane}.} \bibinfo{year}{2023}\natexlab{}.
\newblock \showarticletitle{Artificial hallucinations in ChatGPT: implications in scientific writing}.
\newblock \bibinfo{journal}{\emph{Cureus}} \bibinfo{volume}{15}, \bibinfo{number}{2} (\bibinfo{year}{2023}), \bibinfo{pages}{1--4}.
\newblock
\urldef\tempurl%
\url{https://doi.org/10.7759/cureus.35179}
\showDOI{\tempurl}


\bibitem[Bahrini et~al\mbox{.}(2023)]%
        {10137850}
\bibfield{author}{\bibinfo{person}{Aram Bahrini}, \bibinfo{person}{Mohammadsadra Khamoshifar}, \bibinfo{person}{Hossein Abbasimehr}, \bibinfo{person}{Robert~J. Riggs}, \bibinfo{person}{Maryam Esmaeili}, \bibinfo{person}{Rastin~Mastali Majdabadkohne}, {and} \bibinfo{person}{Morteza Pasehvar}.} \bibinfo{year}{2023}\natexlab{}.
\newblock \showarticletitle{ChatGPT: Applications, Opportunities, and Threats}. In \bibinfo{booktitle}{\emph{2023 Systems and Information Engineering Design Symposium (SIEDS)}}. \bibinfo{publisher}{IEEE}, \bibinfo{address}{Charlottesville, VA, USA}, \bibinfo{pages}{274--279}.
\newblock
\urldef\tempurl%
\url{https://doi.org/10.1109/SIEDS58326.2023.10137850}
\showDOI{\tempurl}


\bibitem[Bakker and Demerouti(2007)]%
        {bakker2007job}
\bibfield{author}{\bibinfo{person}{Arnold~B Bakker} {and} \bibinfo{person}{Evangelia Demerouti}.} \bibinfo{year}{2007}\natexlab{}.
\newblock \showarticletitle{The job demands-resources model: State of the art}.
\newblock \bibinfo{journal}{\emph{Journal of managerial psychology}} \bibinfo{volume}{22}, \bibinfo{number}{3} (\bibinfo{year}{2007}), \bibinfo{pages}{309--328}.
\newblock
\showISSN{0268-3946}
\urldef\tempurl%
\url{https://doi.org/10.1108/02683940710733115}
\showDOI{\tempurl}


\bibitem[{Barredo Arrieta} et~al\mbox{.}(2020)]%
        {BARREDOARRIETA202082}
\bibfield{author}{\bibinfo{person}{Alejandro {Barredo Arrieta}}, \bibinfo{person}{Natalia Díaz-Rodríguez}, \bibinfo{person}{Javier {Del Ser}}, \bibinfo{person}{Adrien Bennetot}, \bibinfo{person}{Siham Tabik}, \bibinfo{person}{Alberto Barbado}, \bibinfo{person}{Salvador Garcia}, \bibinfo{person}{Sergio Gil-Lopez}, \bibinfo{person}{Daniel Molina}, \bibinfo{person}{Richard Benjamins}, \bibinfo{person}{Raja Chatila}, {and} \bibinfo{person}{Francisco Herrera}.} \bibinfo{year}{2020}\natexlab{}.
\newblock \showarticletitle{Explainable Artificial Intelligence (XAI): Concepts, taxonomies, opportunities and challenges toward responsible AI}.
\newblock \bibinfo{journal}{\emph{Information Fusion}}  \bibinfo{volume}{58} (\bibinfo{year}{2020}), \bibinfo{pages}{82--115}.
\newblock
\showISSN{1566-2535}
\urldef\tempurl%
\url{https://doi.org/10.1016/j.inffus.2019.12.012}
\showDOI{\tempurl}


\bibitem[Bazeley(2013)]%
        {bazeley2013qualitative}
\bibfield{author}{\bibinfo{person}{P. Bazeley}.} \bibinfo{year}{2013}\natexlab{}.
\newblock \bibinfo{booktitle}{\emph{Qualitative Data Analysis: Practical Strategies}}.
\newblock \bibinfo{publisher}{SAGE Publications}.
\newblock
\showISBNx{9781446289426}
\urldef\tempurl%
\url{https://books.google.com/books?id=33BEAgAAQBAJ}
\showURL{%
\tempurl}


\bibitem[Bishop(2023)]%
        {bishop2023computer}
\bibfield{author}{\bibinfo{person}{Lea Bishop}.} \bibinfo{year}{2023}\natexlab{}.
\newblock \showarticletitle{A computer wrote this paper: What chatgpt means for education, research, and writing}.
\newblock \bibinfo{journal}{\emph{Research, and Writing (January 26, 2023)}} (\bibinfo{year}{2023}).
\newblock
\urldef\tempurl%
\url{https://doi.org/10.2139/ssrn.4338981}
\showDOI{\tempurl}


\bibitem[Boyatzis(1998)]%
        {boyatzis1998transforming}
\bibfield{author}{\bibinfo{person}{R.E. Boyatzis}.} \bibinfo{year}{1998}\natexlab{}.
\newblock \bibinfo{booktitle}{\emph{Transforming Qualitative Information: Thematic Analysis and Code Development}}.
\newblock \bibinfo{publisher}{SAGE Publications}.
\newblock
\showISBNx{9780761909613}
\showLCCN{97045405}
\urldef\tempurl%
\url{https://books.google.com/books?id=_rfClWRhIKAC}
\showURL{%
\tempurl}


\bibitem[Braun and Clarke(2006)]%
        {braun2006using}
\bibfield{author}{\bibinfo{person}{Virginia Braun} {and} \bibinfo{person}{Victoria Clarke}.} \bibinfo{year}{2006}\natexlab{}.
\newblock \showarticletitle{Using thematic analysis in psychology}.
\newblock \bibinfo{journal}{\emph{Qualitative research in psychology}} \bibinfo{volume}{3}, \bibinfo{number}{2} (\bibinfo{year}{2006}), \bibinfo{pages}{77--101}.
\newblock
\urldef\tempurl%
\url{https://doi.org/10.1191/1478088706qp063oa}
\showDOI{\tempurl}


\bibitem[Braun and Clarke(2012)]%
        {braun2012thematic}
\bibfield{author}{\bibinfo{person}{Virginia Braun} {and} \bibinfo{person}{Victoria Clarke}.} \bibinfo{year}{2012}\natexlab{}.
\newblock \bibinfo{booktitle}{\emph{Thematic analysis.}}
\newblock \bibinfo{publisher}{American Psychological Association}.
\newblock
\urldef\tempurl%
\url{https://doi.org/10.1037/13620-004}
\showDOI{\tempurl}


\bibitem[Braun and Clarke(2019)]%
        {braun2019reflecting}
\bibfield{author}{\bibinfo{person}{Virginia Braun} {and} \bibinfo{person}{Victoria Clarke}.} \bibinfo{year}{2019}\natexlab{}.
\newblock \showarticletitle{Reflecting on reflexive thematic analysis}.
\newblock \bibinfo{journal}{\emph{Qualitative research in sport, exercise and health}} \bibinfo{volume}{11}, \bibinfo{number}{4} (\bibinfo{year}{2019}), \bibinfo{pages}{589--597}.
\newblock
\urldef\tempurl%
\url{https://doi.org/10.1080/2159676X.2019.1628806}
\showDOI{\tempurl}


\bibitem[Brown et~al\mbox{.}(2020)]%
        {NEURIPS2020_1457c0d6}
\bibfield{author}{\bibinfo{person}{Tom Brown}, \bibinfo{person}{Benjamin Mann}, \bibinfo{person}{Nick Ryder}, \bibinfo{person}{Melanie Subbiah}, \bibinfo{person}{Jared~D Kaplan}, \bibinfo{person}{Prafulla Dhariwal}, \bibinfo{person}{Arvind Neelakantan}, \bibinfo{person}{Pranav Shyam}, \bibinfo{person}{Girish Sastry}, \bibinfo{person}{Amanda Askell}, \bibinfo{person}{Sandhini Agarwal}, \bibinfo{person}{Ariel Herbert-Voss}, \bibinfo{person}{Gretchen Krueger}, \bibinfo{person}{Tom Henighan}, \bibinfo{person}{Rewon Child}, \bibinfo{person}{Aditya Ramesh}, \bibinfo{person}{Daniel Ziegler}, \bibinfo{person}{Jeffrey Wu}, \bibinfo{person}{Clemens Winter}, \bibinfo{person}{Chris Hesse}, \bibinfo{person}{Mark Chen}, \bibinfo{person}{Eric Sigler}, \bibinfo{person}{Mateusz Litwin}, \bibinfo{person}{Scott Gray}, \bibinfo{person}{Benjamin Chess}, \bibinfo{person}{Jack Clark}, \bibinfo{person}{Christopher Berner}, \bibinfo{person}{Sam McCandlish}, \bibinfo{person}{Alec Radford}, \bibinfo{person}{Ilya Sutskever}, {and}
  \bibinfo{person}{Dario Amodei}.} \bibinfo{year}{2020}\natexlab{}.
\newblock \showarticletitle{Language Models are Few-Shot Learners}. In \bibinfo{booktitle}{\emph{Advances in Neural Information Processing Systems}}, \bibfield{editor}{\bibinfo{person}{H.~Larochelle}, \bibinfo{person}{M.~Ranzato}, \bibinfo{person}{R.~Hadsell}, \bibinfo{person}{M.F. Balcan}, {and} \bibinfo{person}{H.~Lin}} (Eds.), Vol.~\bibinfo{volume}{33}. \bibinfo{publisher}{Curran Associates, Inc.}, \bibinfo{pages}{1877--1901}.
\newblock
\urldef\tempurl%
\url{https://proceedings.neurips.cc/paper_files/paper/2020/file/1457c0d6bfcb4967418bfb8ac142f64a-Paper.pdf}
\showURL{%
\tempurl}


\bibitem[Bryman(2016)]%
        {bryman2016social}
\bibfield{author}{\bibinfo{person}{A. Bryman}.} \bibinfo{year}{2016}\natexlab{}.
\newblock \bibinfo{booktitle}{\emph{Social Research Methods}}.
\newblock \bibinfo{publisher}{Oxford University Press}.
\newblock
\showISBNx{9780199689453}
\showLCCN{2015940141}
\urldef\tempurl%
\url{https://books.google.com/books?id=N2zQCgAAQBAJ}
\showURL{%
\tempurl}


\bibitem[Buschek et~al\mbox{.}(2022)]%
        {10.1145/3519264}
\bibfield{author}{\bibinfo{person}{Daniel Buschek}, \bibinfo{person}{Malin Eiband}, {and} \bibinfo{person}{Heinrich Hussmann}.} \bibinfo{year}{2022}\natexlab{}.
\newblock \showarticletitle{How to Support Users in Understanding Intelligent Systems? An Analysis and Conceptual Framework of User Questions Considering User Mindsets, Involvement, and Knowledge Outcomes}.
\newblock  (\bibinfo{year}{2022}).
\newblock
\showISSN{2160-6455}
\urldef\tempurl%
\url{https://doi.org/10.1145/3519264}
\showDOI{\tempurl}


\bibitem[Castelvecchi(2016)]%
        {castelvecchi2016can}
\bibfield{author}{\bibinfo{person}{Davide Castelvecchi}.} \bibinfo{year}{2016}\natexlab{}.
\newblock \showarticletitle{Can we open the black box of AI?}
\newblock \bibinfo{journal}{\emph{Nature News}} \bibinfo{volume}{538}, \bibinfo{number}{7623} (\bibinfo{year}{2016}), \bibinfo{pages}{20}.
\newblock
\urldef\tempurl%
\url{https://doi.org/10.1038/538020a}
\showDOI{\tempurl}


\bibitem[Castleberry and Nolen(2018)]%
        {castleberry2018thematic}
\bibfield{author}{\bibinfo{person}{Ashley Castleberry} {and} \bibinfo{person}{Amanda Nolen}.} \bibinfo{year}{2018}\natexlab{}.
\newblock \showarticletitle{Thematic analysis of qualitative research data: Is it as easy as it sounds?}
\newblock \bibinfo{journal}{\emph{Currents in Pharmacy Teaching and Learning}} \bibinfo{volume}{10}, \bibinfo{number}{6} (\bibinfo{year}{2018}), \bibinfo{pages}{807--815}.
\newblock
\showISSN{1877-1297}
\urldef\tempurl%
\url{https://doi.org/10.1016/j.cptl.2018.03.019}
\showDOI{\tempurl}


\bibitem[Chang et~al\mbox{.}(2024)]%
        {10.1145/3641289}
\bibfield{author}{\bibinfo{person}{Yupeng Chang}, \bibinfo{person}{Xu Wang}, \bibinfo{person}{Jindong Wang}, \bibinfo{person}{Yuan Wu}, \bibinfo{person}{Linyi Yang}, \bibinfo{person}{Kaijie Zhu}, \bibinfo{person}{Hao Chen}, \bibinfo{person}{Xiaoyuan Yi}, \bibinfo{person}{Cunxiang Wang}, \bibinfo{person}{Yidong Wang}, \bibinfo{person}{Wei Ye}, \bibinfo{person}{Yue Zhang}, \bibinfo{person}{Yi Chang}, \bibinfo{person}{Philip~S. Yu}, \bibinfo{person}{Qiang Yang}, {and} \bibinfo{person}{Xing Xie}.} \bibinfo{year}{2024}\natexlab{}.
\newblock \showarticletitle{A Survey on Evaluation of Large Language Models}.
\newblock \bibinfo{journal}{\emph{ACM Trans. Intell. Syst. Technol.}} \bibinfo{volume}{15}, \bibinfo{number}{3}, Article \bibinfo{articleno}{39} (\bibinfo{date}{mar} \bibinfo{year}{2024}), \bibinfo{numpages}{45}~pages.
\newblock
\showISSN{2157-6904}
\urldef\tempurl%
\url{https://doi.org/10.1145/3641289}
\showDOI{\tempurl}


\bibitem[Chen et~al\mbox{.}(2018)]%
        {chen2018using}
\bibfield{author}{\bibinfo{person}{Nan-Chen Chen}, \bibinfo{person}{Margaret Drouhard}, \bibinfo{person}{Rafal Kocielnik}, \bibinfo{person}{Jina Suh}, {and} \bibinfo{person}{Cecilia~R Aragon}.} \bibinfo{year}{2018}\natexlab{}.
\newblock \showarticletitle{Using machine learning to support qualitative coding in social science: Shifting the focus to ambiguity}.
\newblock \bibinfo{journal}{\emph{ACM Transactions on Interactive Intelligent Systems (TiiS)}} \bibinfo{volume}{8}, \bibinfo{number}{2} (\bibinfo{year}{2018}), \bibinfo{pages}{1--20}.
\newblock
\urldef\tempurl%
\url{https://doi.org/10.1145/3185515}
\showDOI{\tempurl}


\bibitem[Christou({[n.\,d.]})]%
        {christou2023Iow}
\bibfield{author}{\bibinfo{person}{Prokopis Christou}.} \bibinfo{year}{[n.\,d.]}\natexlab{}.
\newblock \showarticletitle{How to Use Artificial Intelligence (AI) as a Resource, Methodological and Analysis Tool in Qualitative Research?}
\newblock  \bibinfo{volume}{28}, \bibinfo{number}{7} (\bibinfo{year}{[n.\,d.]}), \bibinfo{pages}{1968--1980}.
\newblock
\showISSN{1052-0147}
\urldef\tempurl%
\url{https://doi.org/10.46743/2160-3715/2023.6406}
\showDOI{\tempurl}


\bibitem[Chu et~al\mbox{.}(2016)]%
        {chu2016data}
\bibfield{author}{\bibinfo{person}{Xu Chu}, \bibinfo{person}{Ihab~F. Ilyas}, \bibinfo{person}{Sanjay Krishnan}, {and} \bibinfo{person}{Jiannan Wang}.} \bibinfo{year}{2016}\natexlab{}.
\newblock \showarticletitle{Data Cleaning: Overview and Emerging Challenges}. In \bibinfo{booktitle}{\emph{Proceedings of the 2016 International Conference on Management of Data}} (San Francisco, California, USA) \emph{(\bibinfo{series}{SIGMOD '16})}. \bibinfo{publisher}{Association for Computing Machinery}, \bibinfo{address}{New York, NY, USA}, \bibinfo{pages}{2201–2206}.
\newblock
\showISBNx{9781450335317}
\urldef\tempurl%
\url{https://doi.org/10.1145/2882903.2912574}
\showDOI{\tempurl}


\bibitem[Churchill and Singh({[n.\,d.]})]%
        {churchill2022Evolution}
\bibfield{author}{\bibinfo{person}{Rob Churchill} {and} \bibinfo{person}{Lisa Singh}.} \bibinfo{year}{[n.\,d.]}\natexlab{}.
\newblock \showarticletitle{The {{Evolution}} of {{Topic Modeling}}}.
\newblock   \bibinfo{volume}{54} (\bibinfo{year}{[n.\,d.]}), \bibinfo{pages}{1--35}.
\newblock
Issue 10s.
\showISSN{0360-0300, 1557-7341}
\urldef\tempurl%
\url{https://doi.org/10.1145/3507900}
\showDOI{\tempurl}


\bibitem[Clarke et~al\mbox{.}(2015)]%
        {clarke2015thematic}
\bibfield{author}{\bibinfo{person}{Victoria Clarke}, \bibinfo{person}{Virginia Braun}, {and} \bibinfo{person}{Nikki Hayfield}.} \bibinfo{year}{2015}\natexlab{}.
\newblock \showarticletitle{Thematic analysis}.
\newblock \bibinfo{journal}{\emph{Qualitative psychology: A practical guide to research methods}}  \bibinfo{volume}{3} (\bibinfo{year}{2015}), \bibinfo{pages}{222--248}.
\newblock
\showISBNx{9781473933415}
\urldef\tempurl%
\url{https://doi.org/9781473933415}
\showDOI{\tempurl}


\bibitem[Cooper et~al\mbox{.}(2022)]%
        {10.1145/3491102.3517716}
\bibfield{author}{\bibinfo{person}{Ned Cooper}, \bibinfo{person}{Tiffanie Horne}, \bibinfo{person}{Gillian~R Hayes}, \bibinfo{person}{Courtney Heldreth}, \bibinfo{person}{Michal Lahav}, \bibinfo{person}{Jess Holbrook}, {and} \bibinfo{person}{Lauren Wilcox}.} \bibinfo{year}{2022}\natexlab{}.
\newblock \showarticletitle{A Systematic Review and Thematic Analysis of Community-Collaborative Approaches to Computing Research}. In \bibinfo{booktitle}{\emph{Proceedings of the 2022 CHI Conference on Human Factors in Computing Systems}} (New Orleans, LA, USA) \emph{(\bibinfo{series}{CHI '22})}. \bibinfo{publisher}{Association for Computing Machinery}, \bibinfo{address}{New York, NY, USA}, Article \bibinfo{articleno}{73}, \bibinfo{numpages}{18}~pages.
\newblock
\showISBNx{9781450391573}
\urldef\tempurl%
\url{https://doi.org/10.1145/3491102.3517716}
\showDOI{\tempurl}


\bibitem[Dang et~al\mbox{.}(2023)]%
        {10.1145/3544548.3580969}
\bibfield{author}{\bibinfo{person}{Hai Dang}, \bibinfo{person}{Sven Goller}, \bibinfo{person}{Florian Lehmann}, {and} \bibinfo{person}{Daniel Buschek}.} \bibinfo{year}{2023}\natexlab{}.
\newblock \showarticletitle{Choice Over Control: How Users Write with Large Language Models Using Diegetic and Non-Diegetic Prompting}. In \bibinfo{booktitle}{\emph{Proceedings of the 2023 CHI Conference on Human Factors in Computing Systems}} (Hamburg, Germany) \emph{(\bibinfo{series}{CHI '23})}. \bibinfo{publisher}{Association for Computing Machinery}, \bibinfo{address}{New York, NY, USA}, Article \bibinfo{articleno}{408}, \bibinfo{numpages}{17}~pages.
\newblock
\showISBNx{9781450394215}
\urldef\tempurl%
\url{https://doi.org/10.1145/3544548.3580969}
\showDOI{\tempurl}


\bibitem[Davenport et~al\mbox{.}(2018)]%
        {davenport2018artificial}
\bibfield{author}{\bibinfo{person}{Thomas~H Davenport}, \bibinfo{person}{Rajeev Ronanki}, {et~al\mbox{.}}} \bibinfo{year}{2018}\natexlab{}.
\newblock \showarticletitle{Artificial intelligence for the real world}.
\newblock \bibinfo{journal}{\emph{Harvard business review}} \bibinfo{volume}{96}, \bibinfo{number}{1} (\bibinfo{year}{2018}), \bibinfo{pages}{108--116}.
\newblock
\urldef\tempurl%
\url{https://www.bizjournals.com/boston/news/2018/01/09/hbr-artificial-intelligence-for-the-real-world.html}
\showURL{%
\tempurl}


\bibitem[Dietterich(2019)]%
        {dietterich2019robust}
\bibfield{author}{\bibinfo{person}{Thomas~G Dietterich}.} \bibinfo{year}{2019}\natexlab{}.
\newblock \showarticletitle{Robust artificial intelligence and robust human organizations}.
\newblock \bibinfo{journal}{\emph{Frontiers of Computer Science}}  \bibinfo{volume}{13} (\bibinfo{year}{2019}), \bibinfo{pages}{1--3}.
\newblock
\urldef\tempurl%
\url{https://doi.org/10.1007/s11704-018-8900-4}
\showDOI{\tempurl}


\bibitem[Dong et~al\mbox{.}(2023)]%
        {dong2023self}
\bibfield{author}{\bibinfo{person}{Yihong Dong}, \bibinfo{person}{Xue Jiang}, \bibinfo{person}{Zhi Jin}, {and} \bibinfo{person}{Ge Li}.} \bibinfo{year}{2023}\natexlab{}.
\newblock \bibinfo{title}{Self-collaboration Code Generation via ChatGPT}.
\newblock
\newblock
\showeprint[arxiv]{2304.07590}~[cs.SE]


\bibitem[Dwivedi et~al\mbox{.}(2021)]%
        {dwivedi2021artificial}
\bibfield{author}{\bibinfo{person}{Yogesh~K. Dwivedi}, \bibinfo{person}{Laurie Hughes}, \bibinfo{person}{Elvira Ismagilova}, \bibinfo{person}{Gert Aarts}, \bibinfo{person}{Crispin Coombs}, \bibinfo{person}{Tom Crick}, \bibinfo{person}{Yanqing Duan}, \bibinfo{person}{Rohita Dwivedi}, \bibinfo{person}{John Edwards}, \bibinfo{person}{Aled Eirug}, \bibinfo{person}{Vassilis Galanos}, \bibinfo{person}{P.~Vigneswara Ilavarasan}, \bibinfo{person}{Marijn Janssen}, \bibinfo{person}{Paul Jones}, \bibinfo{person}{Arpan~Kumar Kar}, \bibinfo{person}{Hatice Kizgin}, \bibinfo{person}{Bianca Kronemann}, \bibinfo{person}{Banita Lal}, \bibinfo{person}{Biagio Lucini}, \bibinfo{person}{Rony Medaglia}, \bibinfo{person}{Kenneth {Le Meunier-FitzHugh}}, \bibinfo{person}{Leslie~Caroline {Le Meunier-FitzHugh}}, \bibinfo{person}{Santosh Misra}, \bibinfo{person}{Emmanuel Mogaji}, \bibinfo{person}{Sujeet~Kumar Sharma}, \bibinfo{person}{Jang~Bahadur Singh}, \bibinfo{person}{Vishnupriya Raghavan}, \bibinfo{person}{Ramakrishnan Raman},
  \bibinfo{person}{Nripendra~P. Rana}, \bibinfo{person}{Spyridon Samothrakis}, \bibinfo{person}{Jak Spencer}, \bibinfo{person}{Kuttimani Tamilmani}, \bibinfo{person}{Annie Tubadji}, \bibinfo{person}{Paul Walton}, {and} \bibinfo{person}{Michael~D. Williams}.} \bibinfo{year}{2021}\natexlab{}.
\newblock \showarticletitle{Artificial Intelligence (AI): Multidisciplinary perspectives on emerging challenges, opportunities, and agenda for research, practice and policy}.
\newblock \bibinfo{journal}{\emph{International Journal of Information Management}}  \bibinfo{volume}{57} (\bibinfo{year}{2021}), \bibinfo{pages}{101994}.
\newblock
\showISSN{0268-4012}
\urldef\tempurl%
\url{https://doi.org/10.1016/j.ijinfomgt.2019.08.002}
\showDOI{\tempurl}


\bibitem[Dwivedi et~al\mbox{.}(2023)]%
        {DWIVEDI2023102642}
\bibfield{author}{\bibinfo{person}{Yogesh~K. Dwivedi}, \bibinfo{person}{Nir Kshetri}, \bibinfo{person}{Laurie Hughes}, \bibinfo{person}{Emma~Louise Slade}, \bibinfo{person}{Anand Jeyaraj}, \bibinfo{person}{Arpan~Kumar Kar}, \bibinfo{person}{Abdullah~M. Baabdullah}, \bibinfo{person}{Alex Koohang}, \bibinfo{person}{Vishnupriya Raghavan}, \bibinfo{person}{Manju Ahuja}, \bibinfo{person}{Hanaa Albanna}, \bibinfo{person}{Mousa~Ahmad Albashrawi}, \bibinfo{person}{Adil~S. Al-Busaidi}, \bibinfo{person}{Janarthanan Balakrishnan}, \bibinfo{person}{Yves Barlette}, \bibinfo{person}{Sriparna Basu}, \bibinfo{person}{Indranil Bose}, \bibinfo{person}{Laurence Brooks}, \bibinfo{person}{Dimitrios Buhalis}, \bibinfo{person}{Lemuria Carter}, \bibinfo{person}{Soumyadeb Chowdhury}, \bibinfo{person}{Tom Crick}, \bibinfo{person}{Scott~W. Cunningham}, \bibinfo{person}{Gareth~H. Davies}, \bibinfo{person}{Robert~M. Davison}, \bibinfo{person}{Rahul Dé}, \bibinfo{person}{Denis Dennehy}, \bibinfo{person}{Yanqing Duan},
  \bibinfo{person}{Rameshwar Dubey}, \bibinfo{person}{Rohita Dwivedi}, \bibinfo{person}{John~S. Edwards}, \bibinfo{person}{Carlos Flavián}, \bibinfo{person}{Robin Gauld}, \bibinfo{person}{Varun Grover}, \bibinfo{person}{Mei-Chih Hu}, \bibinfo{person}{Marijn Janssen}, \bibinfo{person}{Paul Jones}, \bibinfo{person}{Iris Junglas}, \bibinfo{person}{Sangeeta Khorana}, \bibinfo{person}{Sascha Kraus}, \bibinfo{person}{Kai~R. Larsen}, \bibinfo{person}{Paul Latreille}, \bibinfo{person}{Sven Laumer}, \bibinfo{person}{F.~Tegwen Malik}, \bibinfo{person}{Abbas Mardani}, \bibinfo{person}{Marcello Mariani}, \bibinfo{person}{Sunil Mithas}, \bibinfo{person}{Emmanuel Mogaji}, \bibinfo{person}{Jeretta~Horn Nord}, \bibinfo{person}{Siobhan O’Connor}, \bibinfo{person}{Fevzi Okumus}, \bibinfo{person}{Margherita Pagani}, \bibinfo{person}{Neeraj Pandey}, \bibinfo{person}{Savvas Papagiannidis}, \bibinfo{person}{Ilias~O. Pappas}, \bibinfo{person}{Nishith Pathak}, \bibinfo{person}{Jan Pries-Heje}, \bibinfo{person}{Ramakrishnan
  Raman}, \bibinfo{person}{Nripendra~P. Rana}, \bibinfo{person}{Sven-Volker Rehm}, \bibinfo{person}{Samuel Ribeiro-Navarrete}, \bibinfo{person}{Alexander Richter}, \bibinfo{person}{Frantz Rowe}, \bibinfo{person}{Suprateek Sarker}, \bibinfo{person}{Bernd~Carsten Stahl}, \bibinfo{person}{Manoj~Kumar Tiwari}, \bibinfo{person}{Wil {van der Aalst}}, \bibinfo{person}{Viswanath Venkatesh}, \bibinfo{person}{Giampaolo Viglia}, \bibinfo{person}{Michael Wade}, \bibinfo{person}{Paul Walton}, \bibinfo{person}{Jochen Wirtz}, {and} \bibinfo{person}{Ryan Wright}.} \bibinfo{year}{2023}\natexlab{}.
\newblock \showarticletitle{Opinion Paper: “So what if ChatGPT wrote it?” Multidisciplinary perspectives on opportunities, challenges and implications of generative conversational AI for research, practice and policy}.
\newblock \bibinfo{journal}{\emph{International Journal of Information Management}}  \bibinfo{volume}{71} (\bibinfo{year}{2023}), \bibinfo{pages}{102642}.
\newblock
\showISSN{0268-4012}
\urldef\tempurl%
\url{https://doi.org/10.1016/j.ijinfomgt.2023.102642}
\showDOI{\tempurl}


\bibitem[Eager and Brunton(2023)]%
        {eager2023prompting}
\bibfield{author}{\bibinfo{person}{Bronwyn Eager} {and} \bibinfo{person}{Ryan Brunton}.} \bibinfo{year}{2023}\natexlab{}.
\newblock \showarticletitle{Prompting higher education towards AI-augmented teaching and learning practice}.
\newblock \bibinfo{journal}{\emph{Journal of University Teaching \& Learning Practice}} \bibinfo{volume}{20}, \bibinfo{number}{5} (\bibinfo{year}{2023}), \bibinfo{pages}{02}.
\newblock
\urldef\tempurl%
\url{https://doi.org/10.53761/1.20.5.02}
\showDOI{\tempurl}


\bibitem[Feuston and Brubaker(2021)]%
        {10.1145/3479856}
\bibfield{author}{\bibinfo{person}{Jessica~L. Feuston} {and} \bibinfo{person}{Jed~R. Brubaker}.} \bibinfo{year}{2021}\natexlab{}.
\newblock \showarticletitle{Putting Tools in Their Place: The Role of Time and Perspective in Human-AI Collaboration for Qualitative Analysis}.
\newblock \bibinfo{journal}{\emph{Proc. ACM Hum.-Comput. Interact.}} \bibinfo{volume}{5}, \bibinfo{number}{CSCW2}, Article \bibinfo{articleno}{469} (\bibinfo{date}{oct} \bibinfo{year}{2021}), \bibinfo{numpages}{25}~pages.
\newblock
\urldef\tempurl%
\url{https://doi.org/10.1145/3479856}
\showDOI{\tempurl}


\bibitem[Fiannaca et~al\mbox{.}({[n.\,d.]})]%
        {fiannaca2023Programming}
\bibfield{author}{\bibinfo{person}{Alexander~J. Fiannaca}, \bibinfo{person}{Chinmay Kulkarni}, \bibinfo{person}{Carrie~J Cai}, {and} \bibinfo{person}{Michael Terry}.} \bibinfo{year}{[n.\,d.]}\natexlab{}.
\newblock \showarticletitle{Programming without a {{Programming Language}}: {{Challenges}} and {{Opportunities}} for {{Designing Developer Tools}} for {{Prompt Programming}}}. In \bibinfo{booktitle}{\emph{Extended {{Abstracts}} of the 2023 {{CHI Conference}} on {{Human Factors}} in {{Computing Systems}}}} ({Hamburg Germany}, 2023-04-19). \bibinfo{publisher}{{ACM}}, \bibinfo{pages}{1--7}.
\newblock
\showISBNx{978-1-4503-9422-2}
\urldef\tempurl%
\url{https://doi.org/10.1145/3544549.3585737}
\showDOI{\tempurl}


\bibitem[Foucault et~al\mbox{.}(1988)]%
        {foucault1988technologies}
\bibfield{author}{\bibinfo{person}{Michel Foucault} {et~al\mbox{.}}} \bibinfo{year}{1988}\natexlab{}.
\newblock \showarticletitle{Technologies of the self}. In \bibinfo{booktitle}{\emph{Technologies of the self: A seminar with Michel Foucault}}, Vol.~\bibinfo{volume}{18}. Amherst.
\newblock
\urldef\tempurl%
\url{https://works.raqsmediacollective.net/wp-content/uploads/2023/04/sarai_reader_03_shaping_technologies.pdf#page=105}
\showURL{%
\tempurl}


\bibitem[Franzoni(2023)]%
        {franzoni2023black}
\bibfield{author}{\bibinfo{person}{Valentina Franzoni}.} \bibinfo{year}{2023}\natexlab{}.
\newblock \showarticletitle{From black box to glass box: advancing transparency in artificial intelligence systems for ethical and trustworthy AI}. In \bibinfo{booktitle}{\emph{Computational Science and Its Applications -- ICCSA 2023 Workshops}}, \bibfield{editor}{\bibinfo{person}{Osvaldo Gervasi}, \bibinfo{person}{Beniamino Murgante}, \bibinfo{person}{Ana Maria A.~C. Rocha}, \bibinfo{person}{Chiara Garau}, \bibinfo{person}{Francesco Scorza}, \bibinfo{person}{Yeliz Karaca}, {and} \bibinfo{person}{Carmelo~M. Torre}} (Eds.). \bibinfo{publisher}{Springer Nature Switzerland}, \bibinfo{address}{Cham}, \bibinfo{pages}{118--130}.
\newblock
\showISBNx{978-3-031-37114-1}
\urldef\tempurl%
\url{https://doi.org/10.1007/978-3-031-37114-1_9}
\showDOI{\tempurl}


\bibitem[Gao({[n.\,d.]})]%
        {gao2023Prompt}
\bibfield{author}{\bibinfo{person}{Andrew Gao}.} \bibinfo{year}{[n.\,d.]}\natexlab{}.
\newblock \bibinfo{booktitle}{\emph{Prompt {{Engineering}} for {{Large Language Models}}}}.
\newblock
\urldef\tempurl%
\url{https://doi.org/10.2139/ssrn.4504303}
\showDOI{\tempurl}


\bibitem[Gao et~al\mbox{.}(2023)]%
        {gao2023coaicoder}
\bibfield{author}{\bibinfo{person}{Jie Gao}, \bibinfo{person}{Kenny Tsu~Wei Choo}, \bibinfo{person}{Junming Cao}, \bibinfo{person}{Roy Ka~Wei Lee}, {and} \bibinfo{person}{Simon Perrault}.} \bibinfo{year}{2023}\natexlab{}.
\newblock \bibinfo{title}{CoAIcoder: Examining the Effectiveness of AI-assisted Human-to-Human Collaboration in Qualitative Analysis}.
\newblock
\newblock
\urldef\tempurl%
\url{https://doi.org/10.48550/arXiv.2304.05560}
\showDOI{\tempurl}
\showeprint[arxiv]{2304.05560}~[cs.HC]


\bibitem[Gebreegziabher et~al\mbox{.}(2023)]%
        {10.1145/3544548.3581352}
\bibfield{author}{\bibinfo{person}{Simret~Araya Gebreegziabher}, \bibinfo{person}{Zheng Zhang}, \bibinfo{person}{Xiaohang Tang}, \bibinfo{person}{Yihao Meng}, \bibinfo{person}{Elena~L. Glassman}, {and} \bibinfo{person}{Toby Jia-Jun Li}.} \bibinfo{year}{2023}\natexlab{}.
\newblock \showarticletitle{PaTAT: Human-AI Collaborative Qualitative Coding with Explainable Interactive Rule Synthesis}. In \bibinfo{booktitle}{\emph{Proceedings of the 2023 CHI Conference on Human Factors in Computing Systems}} (Hamburg, Germany) \emph{(\bibinfo{series}{CHI '23})}. \bibinfo{publisher}{Association for Computing Machinery}, \bibinfo{address}{New York, NY, USA}, Article \bibinfo{articleno}{362}, \bibinfo{numpages}{19}~pages.
\newblock
\showISBNx{9781450394215}
\urldef\tempurl%
\url{https://doi.org/10.1145/3544548.3581352}
\showDOI{\tempurl}


\bibitem[Guest et~al\mbox{.}(2011)]%
        {guest2011applied}
\bibfield{author}{\bibinfo{person}{G. Guest}, \bibinfo{person}{K.M. MacQueen}, {and} \bibinfo{person}{E.E. Namey}.} \bibinfo{year}{2011}\natexlab{}.
\newblock \bibinfo{booktitle}{\emph{Applied Thematic Analysis}}.
\newblock \bibinfo{publisher}{SAGE Publications}.
\newblock
\showISBNx{9781544367217}
\showLCCN{2011031161}
\urldef\tempurl%
\url{https://books.google.com/books?id=Hr11DwAAQBAJ}
\showURL{%
\tempurl}


\bibitem[Guest et~al\mbox{.}(2013)]%
        {guest2013collecting}
\bibfield{author}{\bibinfo{person}{G. Guest}, \bibinfo{person}{E.E. Namey}, {and} \bibinfo{person}{M.L. Mitchell}.} \bibinfo{year}{2013}\natexlab{}.
\newblock \bibinfo{booktitle}{\emph{Collecting Qualitative Data: A Field Manual for Applied Research}}.
\newblock \bibinfo{publisher}{SAGE Publications}.
\newblock
\showISBNx{9781412986847}
\showLCCN{2012009523}
\urldef\tempurl%
\url{https://books.google.com/books?id=--3rmWYKtloC}
\showURL{%
\tempurl}


\bibitem[Guetterman et~al\mbox{.}({[n.\,d.]})]%
        {guetterman2018Augmenting}
\bibfield{author}{\bibinfo{person}{Timothy~C. Guetterman}, \bibinfo{person}{Tammy Chang}, \bibinfo{person}{Melissa DeJonckheere}, \bibinfo{person}{Tanmay Basu}, \bibinfo{person}{Elizabeth Scruggs}, {and} \bibinfo{person}{V.~G.~Vinod Vydiswaran}.} \bibinfo{year}{[n.\,d.]}\natexlab{}.
\newblock \showarticletitle{Augmenting {{Qualitative Text Analysis}} with {{Natural Language Processing}}: {{Methodological Study}}}.
\newblock  \bibinfo{volume}{20}, \bibinfo{number}{6} (\bibinfo{year}{[n.\,d.]}), \bibinfo{pages}{e9702}.
\newblock
\urldef\tempurl%
\url{https://doi.org/10.2196/jmir.9702}
\showDOI{\tempurl}


\bibitem[Hassani and Silva(2023)]%
        {hassani2023role}
\bibfield{author}{\bibinfo{person}{Hossein Hassani} {and} \bibinfo{person}{Emmanuel~Sirmal Silva}.} \bibinfo{year}{2023}\natexlab{}.
\newblock \showarticletitle{The role of ChatGPT in data science: how ai-assisted conversational interfaces are revolutionizing the field}.
\newblock \bibinfo{journal}{\emph{Big data and cognitive computing}} \bibinfo{volume}{7}, \bibinfo{number}{2} (\bibinfo{year}{2023}), \bibinfo{pages}{62}.
\newblock
\urldef\tempurl%
\url{https://doi.org/10.3390/bdcc7020062}
\showDOI{\tempurl}


\bibitem[Hassenzahl(2010)]%
        {hassenzahl2010experience}
\bibfield{author}{\bibinfo{person}{M. Hassenzahl}.} \bibinfo{year}{2010}\natexlab{}.
\newblock \bibinfo{booktitle}{\emph{Experience Design: Technology for All the Right Reasons}}.
\newblock \bibinfo{publisher}{Morgan \& Claypool Publishers}.
\newblock
\showISBNx{9781608450480}
\urldef\tempurl%
\url{https://books.google.com/books?id=GYZgAQAAQBAJ}
\showURL{%
\tempurl}


\bibitem[Hernandez-Bocanegra and Ziegler(2023)]%
        {10.1145/3579541}
\bibfield{author}{\bibinfo{person}{Diana~C. Hernandez-Bocanegra} {and} \bibinfo{person}{J\"{u}rgen Ziegler}.} \bibinfo{year}{2023}\natexlab{}.
\newblock \showarticletitle{Explaining Recommendations through Conversations: Dialog Model and the Effects of Interface Type and Degree of Interactivity}.
\newblock  (\bibinfo{year}{2023}).
\newblock
\showISSN{2160-6455}
\urldef\tempurl%
\url{https://doi.org/10.1145/3579541}
\showDOI{\tempurl}


\bibitem[Heston and Khun({[n.\,d.]})]%
        {heston2023Prompt}
\bibfield{author}{\bibinfo{person}{Thomas~F. Heston} {and} \bibinfo{person}{Charya Khun}.} \bibinfo{year}{[n.\,d.]}\natexlab{}.
\newblock \showarticletitle{Prompt {{Engineering}} in {{Medical Education}}}.
\newblock  \bibinfo{volume}{2}, \bibinfo{number}{3} (\bibinfo{year}{[n.\,d.]}), \bibinfo{pages}{198--205}.
\newblock
Issue 3.
\showISSN{2813-141X}
\urldef\tempurl%
\url{https://doi.org/10.3390/ime2030019}
\showDOI{\tempurl}


\bibitem[Hong et~al\mbox{.}(2022)]%
        {10.1145/3526113.3545681}
\bibfield{author}{\bibinfo{person}{Matt-Heun Hong}, \bibinfo{person}{Lauren~A. Marsh}, \bibinfo{person}{Jessica~L. Feuston}, \bibinfo{person}{Janet Ruppert}, \bibinfo{person}{Jed~R. Brubaker}, {and} \bibinfo{person}{Danielle~Albers Szafir}.} \bibinfo{year}{2022}\natexlab{}.
\newblock \showarticletitle{Scholastic: Graphical Human-AI Collaboration for Inductive and Interpretive Text Analysis}. In \bibinfo{booktitle}{\emph{Proceedings of the 35th Annual ACM Symposium on User Interface Software and Technology}} (Bend, OR, USA) \emph{(\bibinfo{series}{UIST '22})}. \bibinfo{publisher}{Association for Computing Machinery}, \bibinfo{address}{New York, NY, USA}, Article \bibinfo{articleno}{30}, \bibinfo{numpages}{12}~pages.
\newblock
\showISBNx{9781450393201}
\urldef\tempurl%
\url{https://doi.org/10.1145/3526113.3545681}
\showDOI{\tempurl}


\bibitem[Husserl and Moran(2012)]%
        {husserl2012ideas}
\bibfield{author}{\bibinfo{person}{Edmund Husserl} {and} \bibinfo{person}{Dermot Moran}.} \bibinfo{year}{2012}\natexlab{}.
\newblock \bibinfo{booktitle}{\emph{Ideas: General introduction to pure phenomenology}}.
\newblock \bibinfo{publisher}{Routledge}.
\newblock
\urldef\tempurl%
\url{https://doi.org/10.4324/9780203120330}
\showDOI{\tempurl}


\bibitem[Ihde(1991)]%
        {ihde1991instrumental}
\bibfield{author}{\bibinfo{person}{D. Ihde}.} \bibinfo{year}{1991}\natexlab{}.
\newblock \bibinfo{booktitle}{\emph{Instrumental Realism: The Interface Between Philosophy of Science and Philosophy of Technology}}.
\newblock \bibinfo{publisher}{Indiana University Press}.
\newblock
\showISBNx{9780253206268}
\showLCCN{90042334}
\urldef\tempurl%
\url{https://books.google.com/books?id=vYl0tagdh1QC}
\showURL{%
\tempurl}


\bibitem[Jacovi et~al\mbox{.}(2021)]%
        {10.1145/3442188.3445923}
\bibfield{author}{\bibinfo{person}{Alon Jacovi}, \bibinfo{person}{Ana Marasovi\'{c}}, \bibinfo{person}{Tim Miller}, {and} \bibinfo{person}{Yoav Goldberg}.} \bibinfo{year}{2021}\natexlab{}.
\newblock \showarticletitle{Formalizing Trust in Artificial Intelligence: Prerequisites, Causes and Goals of Human Trust in AI}. In \bibinfo{booktitle}{\emph{Proceedings of the 2021 ACM Conference on Fairness, Accountability, and Transparency}} (Virtual Event, Canada) \emph{(\bibinfo{series}{FAccT '21})}. \bibinfo{publisher}{Association for Computing Machinery}, \bibinfo{address}{New York, NY, USA}, \bibinfo{pages}{624–635}.
\newblock
\showISBNx{9781450383097}
\urldef\tempurl%
\url{https://doi.org/10.1145/3442188.3445923}
\showDOI{\tempurl}


\bibitem[Jacucci et~al\mbox{.}(2014)]%
        {10.1007/978-3-319-13500-7_1}
\bibfield{author}{\bibinfo{person}{Giulio Jacucci}, \bibinfo{person}{Anna Spagnolli}, \bibinfo{person}{Jonathan Freeman}, {and} \bibinfo{person}{Luciano Gamberini}.} \bibinfo{year}{2014}\natexlab{}.
\newblock \showarticletitle{Symbiotic Interaction: A Critical Definition and Comparison to other Human-Computer Paradigms}. In \bibinfo{booktitle}{\emph{Symbiotic Interaction}}, \bibfield{editor}{\bibinfo{person}{Giulio Jacucci}, \bibinfo{person}{Luciano Gamberini}, \bibinfo{person}{Jonathan Freeman}, {and} \bibinfo{person}{Anna Spagnolli}} (Eds.). \bibinfo{publisher}{Springer International Publishing}, \bibinfo{address}{Cham}, \bibinfo{pages}{3--20}.
\newblock
\showISBNx{978-3-319-13500-7}
\urldef\tempurl%
\url{https://doi.org/10.1007/978-3-319-13500-7_1}
\showDOI{\tempurl}


\bibitem[Jarrahi(2018)]%
        {JARRAHI2018577}
\bibfield{author}{\bibinfo{person}{Mohammad~Hossein Jarrahi}.} \bibinfo{year}{2018}\natexlab{}.
\newblock \showarticletitle{Artificial intelligence and the future of work: Human-AI symbiosis in organizational decision making}.
\newblock \bibinfo{journal}{\emph{Business Horizons}} \bibinfo{volume}{61}, \bibinfo{number}{4} (\bibinfo{year}{2018}), \bibinfo{pages}{577--586}.
\newblock
\showISSN{0007-6813}
\urldef\tempurl%
\url{https://doi.org/10.1016/j.bushor.2018.03.007}
\showDOI{\tempurl}


\bibitem[Jensen(2002)]%
        {jensen2013handbook}
\bibfield{author}{\bibinfo{person}{Klaus~Bruhn Jensen}.} \bibinfo{year}{2002}\natexlab{}.
\newblock \bibinfo{booktitle}{\emph{The qualitative research process} (\bibinfo{edition}{1st} ed.)}.
\newblock \bibinfo{publisher}{Routledge}, \bibinfo{address}{London}, Chapter~14, \bibinfo{pages}{247--265}.
\newblock
\showISBNx{9780203465103}
\urldef\tempurl%
\url{https://doi.org/10.4324/9780203465103}
\showDOI{\tempurl}


\bibitem[Jiang et~al\mbox{.}(2022)]%
        {10.1145/3491102.3501870}
\bibfield{author}{\bibinfo{person}{Ellen Jiang}, \bibinfo{person}{Edwin Toh}, \bibinfo{person}{Alejandra Molina}, \bibinfo{person}{Kristen Olson}, \bibinfo{person}{Claire Kayacik}, \bibinfo{person}{Aaron Donsbach}, \bibinfo{person}{Carrie~J Cai}, {and} \bibinfo{person}{Michael Terry}.} \bibinfo{year}{2022}\natexlab{}.
\newblock \showarticletitle{Discovering the Syntax and Strategies of Natural Language Programming with Generative Language Models}. In \bibinfo{booktitle}{\emph{Proceedings of the 2022 CHI Conference on Human Factors in Computing Systems}} (New Orleans, LA, USA) \emph{(\bibinfo{series}{CHI '22})}. \bibinfo{publisher}{Association for Computing Machinery}, \bibinfo{address}{New York, NY, USA}, Article \bibinfo{articleno}{386}, \bibinfo{numpages}{19}~pages.
\newblock
\showISBNx{9781450391573}
\urldef\tempurl%
\url{https://doi.org/10.1145/3491102.3501870}
\showDOI{\tempurl}


\bibitem[Jiang et~al\mbox{.}(2023)]%
        {jiang2023structgpt}
\bibfield{author}{\bibinfo{person}{Jinhao Jiang}, \bibinfo{person}{Kun Zhou}, \bibinfo{person}{Zican Dong}, \bibinfo{person}{Keming Ye}, \bibinfo{person}{Wayne~Xin Zhao}, {and} \bibinfo{person}{Ji-Rong Wen}.} \bibinfo{year}{2023}\natexlab{}.
\newblock \bibinfo{title}{StructGPT: A General Framework for Large Language Model to Reason over Structured Data}.
\newblock
\newblock
\showeprint[arxiv]{2305.09645}~[cs.CL]


\bibitem[Jiang et~al\mbox{.}(2021)]%
        {10.1145/3449168}
\bibfield{author}{\bibinfo{person}{Jialun~Aaron Jiang}, \bibinfo{person}{Kandrea Wade}, \bibinfo{person}{Casey Fiesler}, {and} \bibinfo{person}{Jed~R. Brubaker}.} \bibinfo{year}{2021}\natexlab{}.
\newblock \showarticletitle{Supporting Serendipity: Opportunities and Challenges for Human-AI Collaboration in Qualitative Analysis}.
\newblock \bibinfo{journal}{\emph{Proc. ACM Hum.-Comput. Interact.}} \bibinfo{volume}{5}, \bibinfo{number}{CSCW1}, Article \bibinfo{articleno}{94} (\bibinfo{date}{apr} \bibinfo{year}{2021}), \bibinfo{numpages}{23}~pages.
\newblock
\urldef\tempurl%
\url{https://doi.org/10.1145/3449168}
\showDOI{\tempurl}


\bibitem[Jiao et~al\mbox{.}(2023)]%
        {jiao2023chatgpt}
\bibfield{author}{\bibinfo{person}{Wenxiang Jiao}, \bibinfo{person}{Wenxuan Wang}, \bibinfo{person}{Jen tse Huang}, \bibinfo{person}{Xing Wang}, {and} \bibinfo{person}{Zhaopeng Tu}.} \bibinfo{year}{2023}\natexlab{}.
\newblock \bibinfo{title}{Is ChatGPT A Good Translator? Yes With GPT-4 As The Engine}.
\newblock
\newblock
\showeprint[arxiv]{2301.08745}~[cs.CL]


\bibitem[Katz et~al\mbox{.}({[n.\,d.]})]%
        {katz2023Exploringa}
\bibfield{author}{\bibinfo{person}{Andrew Katz}, \bibinfo{person}{Siqing Wei}, \bibinfo{person}{Gaurav Nanda}, \bibinfo{person}{Christopher Brinton}, {and} \bibinfo{person}{Matthew Ohland}.} \bibinfo{year}{[n.\,d.]}\natexlab{}.
\newblock \bibinfo{booktitle}{\emph{Exploring the {{Efficacy}} of {{ChatGPT}} in {{Analyzing Student Teamwork Feedback}} with an {{Existing Taxonomy}}}}.
\newblock
\urldef\tempurl%
\url{https://doi.org/10.48550/arXiv.2305.11882}
\showDOI{\tempurl}
\showeprint[arxiv]{2305.11882}~[cs]


\bibitem[Khanbhai et~al\mbox{.}({[n.\,d.]})]%
        {khanbhai2021applying}
\bibfield{author}{\bibinfo{person}{Mustafa Khanbhai}, \bibinfo{person}{Patrick Anyadi}, \bibinfo{person}{Joshua Symons}, \bibinfo{person}{Kelsey Flott}, \bibinfo{person}{Ara Darzi}, {and} \bibinfo{person}{Erik Mayer}.} \bibinfo{year}{[n.\,d.]}\natexlab{}.
\newblock \showarticletitle{Applying Natural Language Processing and Machine Learning Techniques to Patient Experience Feedback: A Systematic Review}.
\newblock  \bibinfo{volume}{28}, \bibinfo{number}{1} (\bibinfo{year}{[n.\,d.]}), \bibinfo{pages}{e100262}.
\newblock
\showISSN{2632-1009}
\urldef\tempurl%
\url{https://doi.org/10.1136/bmjhci-2020-100262}
\showDOI{\tempurl}
\showeprint[pmid]{33653690}


\bibitem[Kilhoffer et~al\mbox{.}(2023)]%
        {kilhoffer2023ai}
\bibfield{author}{\bibinfo{person}{Zachary Kilhoffer}, \bibinfo{person}{Anita Nlkolich}, \bibinfo{person}{Madelyn~Rose Sanfilippo}, {and} \bibinfo{person}{Zhixuan Zhou}.} \bibinfo{year}{2023}\natexlab{}.
\newblock \showarticletitle{AI Accountability Policy}.
\newblock \bibinfo{journal}{\emph{NTIA-2023-0005-0810}} (\bibinfo{year}{2023}).
\newblock
\urldef\tempurl%
\url{https://www.ideals.illinois.edu/items/127130}
\showURL{%
\tempurl}


\bibitem[Koivisto and Grassini(2023)]%
        {koivisto2023best}
\bibfield{author}{\bibinfo{person}{Mika Koivisto} {and} \bibinfo{person}{Simone Grassini}.} \bibinfo{year}{2023}\natexlab{}.
\newblock \showarticletitle{Best humans still outperform artificial intelligence in a creative divergent thinking task}.
\newblock \bibinfo{journal}{\emph{Scientific reports}} \bibinfo{volume}{13}, \bibinfo{number}{1} (\bibinfo{year}{2023}), \bibinfo{pages}{13601}.
\newblock
\urldef\tempurl%
\url{https://doi.org/10.1038/s41598-023-40858-3}
\showDOI{\tempurl}


\bibitem[Kong et~al\mbox{.}(2023)]%
        {kong2023better}
\bibfield{author}{\bibinfo{person}{Aobo Kong}, \bibinfo{person}{Shiwan Zhao}, \bibinfo{person}{Hao Chen}, \bibinfo{person}{Qicheng Li}, \bibinfo{person}{Yong Qin}, \bibinfo{person}{Ruiqi Sun}, {and} \bibinfo{person}{Xin Zhou}.} \bibinfo{year}{2023}\natexlab{}.
\newblock \bibinfo{title}{Better Zero-Shot Reasoning with Role-Play Prompting}.
\newblock
\newblock
\showeprint[arxiv]{2308.07702}~[cs.CL]


\bibitem[Kuang et~al\mbox{.}(2022)]%
        {10.1145/3491102.3517647}
\bibfield{author}{\bibinfo{person}{Emily Kuang}, \bibinfo{person}{Xiaofu Jin}, {and} \bibinfo{person}{Mingming Fan}.} \bibinfo{year}{2022}\natexlab{}.
\newblock \showarticletitle{“Merging Results Is No Easy Task”: An International Survey Study of Collaborative Data Analysis Practices Among UX Practitioners}. In \bibinfo{booktitle}{\emph{Proceedings of the 2022 CHI Conference on Human Factors in Computing Systems}} (New Orleans, LA, USA) \emph{(\bibinfo{series}{CHI '22})}. \bibinfo{publisher}{Association for Computing Machinery}, \bibinfo{address}{New York, NY, USA}, Article \bibinfo{articleno}{318}, \bibinfo{numpages}{16}~pages.
\newblock
\showISBNx{9781450391573}
\urldef\tempurl%
\url{https://doi.org/10.1145/3491102.3517647}
\showDOI{\tempurl}


\bibitem[Leung(2015)]%
        {leung2015validity}
\bibfield{author}{\bibinfo{person}{Lawrence Leung}.} \bibinfo{year}{2015}\natexlab{}.
\newblock \showarticletitle{Validity, reliability, and generalizability in qualitative research}.
\newblock \bibinfo{journal}{\emph{Journal of family medicine and primary care}} \bibinfo{volume}{4}, \bibinfo{number}{3} (\bibinfo{year}{2015}), \bibinfo{pages}{324}.
\newblock
\urldef\tempurl%
\url{https://doi.org/10.4103/2249-4863.161306}
\showDOI{\tempurl}


\bibitem[Levac et~al\mbox{.}(2019)]%
        {levac2019scoping}
\bibfield{author}{\bibinfo{person}{Leah Levac}, \bibinfo{person}{Scott Ronis}, \bibinfo{person}{Yuriko Cowper-Smith}, {and} \bibinfo{person}{Oriana Vaccarino}.} \bibinfo{year}{2019}\natexlab{}.
\newblock \showarticletitle{A scoping review: The utility of participatory research approaches in psychology}.
\newblock \bibinfo{journal}{\emph{Journal of Community Psychology}} \bibinfo{volume}{47}, \bibinfo{number}{8} (\bibinfo{year}{2019}), \bibinfo{pages}{1865--1892}.
\newblock
\urldef\tempurl%
\url{https://doi.org/10.1002/jcop.22231}
\showDOI{\tempurl}


\bibitem[Liebrenz et~al\mbox{.}(2023)]%
        {liebrenz2023generating}
\bibfield{author}{\bibinfo{person}{Michael Liebrenz}, \bibinfo{person}{Roman Schleifer}, \bibinfo{person}{Anna Buadze}, \bibinfo{person}{Dinesh Bhugra}, {and} \bibinfo{person}{Alexander Smith}.} \bibinfo{year}{2023}\natexlab{}.
\newblock \showarticletitle{Generating scholarly content with ChatGPT: ethical challenges for medical publishing}.
\newblock \bibinfo{journal}{\emph{The Lancet Digital Health}} \bibinfo{volume}{5}, \bibinfo{number}{3} (\bibinfo{year}{2023}), \bibinfo{pages}{e105--e106}.
\newblock
\urldef\tempurl%
\url{https://doi.org/10.1016/S2589-7500(23)00019-5}
\showDOI{\tempurl}


\bibitem[Linzbach et~al\mbox{.}(2023)]%
        {10.1145/3543873.3587655}
\bibfield{author}{\bibinfo{person}{Stephan Linzbach}, \bibinfo{person}{Tim Tressel}, \bibinfo{person}{Laura Kallmeyer}, \bibinfo{person}{Stefan Dietze}, {and} \bibinfo{person}{Hajira Jabeen}.} \bibinfo{year}{2023}\natexlab{}.
\newblock \showarticletitle{Decoding Prompt Syntax: Analysing Its Impact on Knowledge Retrieval in Large Language Models}. In \bibinfo{booktitle}{\emph{Companion Proceedings of the ACM Web Conference 2023}} (Austin, TX, USA) \emph{(\bibinfo{series}{WWW '23 Companion})}. \bibinfo{publisher}{Association for Computing Machinery}, \bibinfo{address}{New York, NY, USA}, \bibinfo{pages}{1145–1149}.
\newblock
\showISBNx{9781450394192}
\urldef\tempurl%
\url{https://doi.org/10.1145/3543873.3587655}
\showDOI{\tempurl}


\bibitem[Liu et~al\mbox{.}(2023)]%
        {liu2023jailbreaking}
\bibfield{author}{\bibinfo{person}{Yi Liu}, \bibinfo{person}{Gelei Deng}, \bibinfo{person}{Zhengzi Xu}, \bibinfo{person}{Yuekang Li}, \bibinfo{person}{Yaowen Zheng}, \bibinfo{person}{Ying Zhang}, \bibinfo{person}{Lida Zhao}, \bibinfo{person}{Tianwei Zhang}, {and} \bibinfo{person}{Yang Liu}.} \bibinfo{year}{2023}\natexlab{}.
\newblock \bibinfo{title}{Jailbreaking ChatGPT via Prompt Engineering: An Empirical Study}.
\newblock
\newblock
\showeprint[arxiv]{2305.13860}~[cs.SE]


\bibitem[Macdonald et~al\mbox{.}(2023)]%
        {macdonald2023can}
\bibfield{author}{\bibinfo{person}{Calum Macdonald}, \bibinfo{person}{Davies Adeloye}, \bibinfo{person}{Aziz Sheikh}, {and} \bibinfo{person}{Igor Rudan}.} \bibinfo{year}{2023}\natexlab{}.
\newblock \showarticletitle{Can ChatGPT draft a research article? An example of population-level vaccine effectiveness analysis}.
\newblock \bibinfo{journal}{\emph{Journal of global health}}  \bibinfo{volume}{13} (\bibinfo{year}{2023}).
\newblock
\urldef\tempurl%
\url{https://doi.org/10.7189/jogh.13.01003}
\showDOI{\tempurl}


\bibitem[Maguire and Delahunt(2017)]%
        {maguire2017doing}
\bibfield{author}{\bibinfo{person}{Moira Maguire} {and} \bibinfo{person}{Brid Delahunt}.} \bibinfo{year}{2017}\natexlab{}.
\newblock \showarticletitle{Doing a thematic analysis: A practical, step-by-step guide for learning and teaching scholars.}
\newblock \bibinfo{journal}{\emph{All Ireland Journal of Higher Education}} \bibinfo{volume}{9}, \bibinfo{number}{3} (\bibinfo{year}{2017}).
\newblock
\urldef\tempurl%
\url{https://ojs.aishe.org/index.php/aishe-j/article/view/335/553}
\showURL{%
\tempurl}


\bibitem[Marcoci et~al\mbox{.}(2023)]%
        {marcoci2023big}
\bibfield{author}{\bibinfo{person}{Alexandru Marcoci}, \bibinfo{person}{Ann~C Thresher}, \bibinfo{person}{Niels~CM Martens}, \bibinfo{person}{Peter Galison}, \bibinfo{person}{Sheperd~S Doeleman}, {and} \bibinfo{person}{Michael~D Johnson}.} \bibinfo{year}{2023}\natexlab{}.
\newblock \showarticletitle{Big STEM collaborations should include humanities and social science}.
\newblock \bibinfo{journal}{\emph{Nature Human Behaviour}} (\bibinfo{year}{2023}), \bibinfo{pages}{1--2}.
\newblock
\urldef\tempurl%
\url{https://doi.org/10.1038/s41562-023-01674-x}
\showDOI{\tempurl}


\bibitem[Megahed et~al\mbox{.}(2023)]%
        {doi:10.1080/08982112.2023.2206479}
\bibfield{author}{\bibinfo{person}{Fadel~M. Megahed}, \bibinfo{person}{Ying-Ju Chen}, \bibinfo{person}{Joshua~A. Ferris}, \bibinfo{person}{Sven Knoth}, {and} \bibinfo{person}{L.~Allison Jones-Farmer}.} \bibinfo{year}{2023}\natexlab{}.
\newblock \showarticletitle{How generative AI models such as ChatGPT can be (mis)used in SPC practice, education, and research? An exploratory study}.
\newblock \bibinfo{journal}{\emph{Quality Engineering}} \bibinfo{volume}{0}, \bibinfo{number}{0} (\bibinfo{year}{2023}), \bibinfo{pages}{1--29}.
\newblock
\urldef\tempurl%
\url{https://doi.org/10.1080/08982112.2023.2206479}
\showDOI{\tempurl}
\showeprint{https://doi.org/10.1080/08982112.2023.2206479}


\bibitem[Mohseni et~al\mbox{.}(2021)]%
        {10.1145/3387166}
\bibfield{author}{\bibinfo{person}{Sina Mohseni}, \bibinfo{person}{Niloofar Zarei}, {and} \bibinfo{person}{Eric~D. Ragan}.} \bibinfo{year}{2021}\natexlab{}.
\newblock \showarticletitle{A Multidisciplinary Survey and Framework for Design and Evaluation of Explainable AI Systems}.
\newblock  (\bibinfo{year}{2021}).
\newblock
\showISSN{2160-6455}
\urldef\tempurl%
\url{https://doi.org/10.1145/3387166}
\showDOI{\tempurl}


\bibitem[Morgan(2022)]%
        {morgan2022understanding}
\bibfield{author}{\bibinfo{person}{Hani Morgan}.} \bibinfo{year}{2022}\natexlab{}.
\newblock \showarticletitle{Understanding thematic analysis and the debates involving its use}.
\newblock \bibinfo{journal}{\emph{The Qualitative Report}} \bibinfo{volume}{27}, \bibinfo{number}{10} (\bibinfo{year}{2022}), \bibinfo{pages}{2079--2090}.
\newblock
\urldef\tempurl%
\url{https://doi.org/10.46743/2160-3715/2022.5912}
\showDOI{\tempurl}


\bibitem[Motoki et~al\mbox{.}(2023)]%
        {motoki2023more}
\bibfield{author}{\bibinfo{person}{Fabio Motoki}, \bibinfo{person}{Valdemar Pinho~Neto}, {and} \bibinfo{person}{Victor Rodrigues}.} \bibinfo{year}{2023}\natexlab{}.
\newblock \showarticletitle{More human than human: Measuring chatgpt political bias}.
\newblock \bibinfo{journal}{\emph{Available at SSRN 4372349}} (\bibinfo{year}{2023}).
\newblock
\urldef\tempurl%
\url{https://doi.org/10.2139/ssrn.4372349}
\showDOI{\tempurl}


\bibitem[Nissenbaum(2004)]%
        {nissenbaum2010privacy}
\bibfield{author}{\bibinfo{person}{Helen Nissenbaum}.} \bibinfo{year}{2004}\natexlab{}.
\newblock \showarticletitle{Privacy as Contextual Integrity Symposium: Technology, Values, and the Justice System}.
\newblock \bibinfo{journal}{\emph{Washington Law Review}}  \bibinfo{volume}{79} (\bibinfo{year}{2004}), \bibinfo{pages}{119}.
\newblock
\urldef\tempurl%
\url{https://doi.org/HOL/Page?handle=hein.journals/washlr79&id=129}
\showDOI{\tempurl}


\bibitem[Nowell et~al\mbox{.}(2017)]%
        {nowell2017thematic}
\bibfield{author}{\bibinfo{person}{Lorelli~S Nowell}, \bibinfo{person}{Jill~M Norris}, \bibinfo{person}{Deborah~E White}, {and} \bibinfo{person}{Nancy~J Moules}.} \bibinfo{year}{2017}\natexlab{}.
\newblock \showarticletitle{Thematic analysis: Striving to meet the trustworthiness criteria}.
\newblock \bibinfo{journal}{\emph{International journal of qualitative methods}} \bibinfo{volume}{16}, \bibinfo{number}{1} (\bibinfo{year}{2017}), \bibinfo{pages}{1609406917733847}.
\newblock
\urldef\tempurl%
\url{https://doi.org/10.1177/1609406917733847}
\showDOI{\tempurl}


\bibitem[Ortloff et~al\mbox{.}(2023)]%
        {10.1145/3544548.3580766}
\bibfield{author}{\bibinfo{person}{Anna-Marie Ortloff}, \bibinfo{person}{Matthias Fassl}, \bibinfo{person}{Alexander Ponticello}, \bibinfo{person}{Florin Martius}, \bibinfo{person}{Anne Mertens}, \bibinfo{person}{Katharina Krombholz}, {and} \bibinfo{person}{Matthew Smith}.} \bibinfo{year}{2023}\natexlab{}.
\newblock \showarticletitle{Different Researchers, Different Results? Analyzing the Influence of Researcher Experience and Data Type During Qualitative Analysis of an Interview and Survey Study on Security Advice}. In \bibinfo{booktitle}{\emph{Proceedings of the 2023 CHI Conference on Human Factors in Computing Systems}} (Hamburg, Germany) \emph{(\bibinfo{series}{CHI '23})}. \bibinfo{publisher}{Association for Computing Machinery}, \bibinfo{address}{New York, NY, USA}, Article \bibinfo{articleno}{864}, \bibinfo{numpages}{21}~pages.
\newblock
\showISBNx{9781450394215}
\urldef\tempurl%
\url{https://doi.org/10.1145/3544548.3580766}
\showDOI{\tempurl}


\bibitem[Parameswaran et~al\mbox{.}(2020)]%
        {parameswaran2020live}
\bibfield{author}{\bibinfo{person}{Uma~D Parameswaran}, \bibinfo{person}{Jade~L Ozawa-Kirk}, {and} \bibinfo{person}{Gwen Latendresse}.} \bibinfo{year}{2020}\natexlab{}.
\newblock \showarticletitle{To live (code) or to not: A new method for coding in qualitative research}.
\newblock \bibinfo{journal}{\emph{Qualitative social work}} \bibinfo{volume}{19}, \bibinfo{number}{4} (\bibinfo{year}{2020}), \bibinfo{pages}{630--644}.
\newblock
\urldef\tempurl%
\url{https://doi.org/10.1177/1473325019840394}
\showDOI{\tempurl}


\bibitem[Paul and Elder(2004)]%
        {paul2004critical}
\bibfield{author}{\bibinfo{person}{Richard Paul} {and} \bibinfo{person}{Linda Elder}.} \bibinfo{year}{2004}\natexlab{}.
\newblock \showarticletitle{Critical and creative thinking}.
\newblock \bibinfo{journal}{\emph{Dillon Beach, CA: The Foundation for Critical Thinking}} (\bibinfo{year}{2004}).
\newblock
\showISBNx{978-0-944583-26-5}
\urldef\tempurl%
\url{https://www.thefont.co.za/wp-content/uploads/2020/03/007-Paul-Elder-2008-A-FULL-Guide-to-Crit-Thinking-Concepts_Tool.pdf}
\showURL{%
\tempurl}


\bibitem[Radford et~al\mbox{.}(2019)]%
        {radford2019language}
\bibfield{author}{\bibinfo{person}{Alec Radford}, \bibinfo{person}{Jeffrey Wu}, \bibinfo{person}{Rewon Child}, \bibinfo{person}{David Luan}, \bibinfo{person}{Dario Amodei}, \bibinfo{person}{Ilya Sutskever}, {et~al\mbox{.}}} \bibinfo{year}{2019}\natexlab{}.
\newblock \showarticletitle{Language models are unsupervised multitask learners}.
\newblock \bibinfo{journal}{\emph{OpenAI blog}} \bibinfo{volume}{1}, \bibinfo{number}{8} (\bibinfo{year}{2019}), \bibinfo{pages}{9}.
\newblock
\urldef\tempurl%
\url{https://api.semanticscholar.org/CorpusID:160025533}
\showURL{%
\tempurl}


\bibitem[Rainey et~al\mbox{.}(2022)]%
        {10.1145/3491102.3502103}
\bibfield{author}{\bibinfo{person}{Jay Rainey}, \bibinfo{person}{Siobhan Macfarlane}, \bibinfo{person}{Aare Puussaar}, \bibinfo{person}{Vasilis Vlachokyriakos}, \bibinfo{person}{Roger Burrows}, \bibinfo{person}{Jan~David Smeddinck}, \bibinfo{person}{Pamela Briggs}, {and} \bibinfo{person}{Kyle Montague}.} \bibinfo{year}{2022}\natexlab{}.
\newblock \showarticletitle{Exploring the Role of Paradata in Digitally Supported Qualitative Co-Research}. In \bibinfo{booktitle}{\emph{Proceedings of the 2022 CHI Conference on Human Factors in Computing Systems}} (New Orleans, LA, USA) \emph{(\bibinfo{series}{CHI '22})}. \bibinfo{publisher}{Association for Computing Machinery}, \bibinfo{address}{New York, NY, USA}, Article \bibinfo{articleno}{584}, \bibinfo{numpages}{16}~pages.
\newblock
\showISBNx{9781450391573}
\urldef\tempurl%
\url{https://doi.org/10.1145/3491102.3502103}
\showDOI{\tempurl}


\bibitem[Richards and Hemphill(2018)]%
        {richards2018practical}
\bibfield{author}{\bibinfo{person}{K~Andrew~R Richards} {and} \bibinfo{person}{Michael~A Hemphill}.} \bibinfo{year}{2018}\natexlab{}.
\newblock \showarticletitle{A practical guide to collaborative qualitative data analysis}.
\newblock \bibinfo{journal}{\emph{Journal of Teaching in Physical education}} \bibinfo{volume}{37}, \bibinfo{number}{2} (\bibinfo{year}{2018}), \bibinfo{pages}{225--231}.
\newblock
\urldef\tempurl%
\url{https://doi.org/10.1123/jtpe.2017-0084}
\showDOI{\tempurl}


\bibitem[Roumeliotis and Tselikas(2023)]%
        {roumeliotis2023chatgpt}
\bibfield{author}{\bibinfo{person}{Konstantinos~I Roumeliotis} {and} \bibinfo{person}{Nikolaos~D Tselikas}.} \bibinfo{year}{2023}\natexlab{}.
\newblock \showarticletitle{ChatGPT and Open-AI Models: A Preliminary Review}.
\newblock \bibinfo{journal}{\emph{Future Internet}} \bibinfo{volume}{15}, \bibinfo{number}{6} (\bibinfo{year}{2023}), \bibinfo{pages}{192}.
\newblock
\urldef\tempurl%
\url{https://doi.org/10.3390/fi15060192}
\showDOI{\tempurl}


\bibitem[Schrills and Franke(2023)]%
        {10.1145/3588594}
\bibfield{author}{\bibinfo{person}{Tim Schrills} {and} \bibinfo{person}{Thomas Franke}.} \bibinfo{year}{2023}\natexlab{}.
\newblock \showarticletitle{How Do Users Experience Traceability of AI Systems? Examining Subjective Information Processing Awareness in Automated Insulin Delivery (AID) Systems}.
\newblock  \bibinfo{volume}{13}, \bibinfo{number}{4} (\bibinfo{year}{2023}).
\newblock
\showISSN{2160-6455}
\urldef\tempurl%
\url{https://doi.org/10.1145/3588594}
\showDOI{\tempurl}


\bibitem[Shen et~al\mbox{.}(2023)]%
        {shen2023chatgpt}
\bibfield{author}{\bibinfo{person}{Yiqiu Shen}, \bibinfo{person}{Laura Heacock}, \bibinfo{person}{Jonathan Elias}, \bibinfo{person}{Keith~D Hentel}, \bibinfo{person}{Beatriu Reig}, \bibinfo{person}{George Shih}, {and} \bibinfo{person}{Linda Moy}.} \bibinfo{year}{2023}\natexlab{}.
\newblock \bibinfo{title}{ChatGPT and other large language models are double-edged swords}.
\newblock , \bibinfo{numpages}{e230163}~pages.
\newblock
\urldef\tempurl%
\url{https://doi.org/10.1148/radiol.230163}
\showDOI{\tempurl}


\bibitem[Siiman et~al\mbox{.}({[n.\,d.]})]%
        {siiman2023Opportunities}
\bibfield{author}{\bibinfo{person}{Leo~A. Siiman}, \bibinfo{person}{Meeli Rannastu-Avalos}, \bibinfo{person}{Johanna Pöysä-Tarhonen}, \bibinfo{person}{Päivi Häkkinen}, {and} \bibinfo{person}{Margus Pedaste}.} \bibinfo{year}{[n.\,d.]}\natexlab{}.
\newblock \showarticletitle{Opportunities and {{Challenges}} for {{AI-Assisted Qualitative Data Analysis}}: {{An Example}} from {{Collaborative Problem-Solving Discourse Data}}}. In \bibinfo{booktitle}{\emph{Innovative {{Technologies}} and {{Learning}}}} ({Cham}, 2023) \emph{(\bibinfo{series}{Lecture {{Notes}} in {{Computer Science}}})}, \bibfield{editor}{\bibinfo{person}{Yueh-Min Huang} {and} \bibinfo{person}{Tânia Rocha}} (Eds.). \bibinfo{publisher}{{Springer Nature Switzerland}}, \bibinfo{pages}{87--96}.
\newblock
\showISBNx{978-3-031-40113-8}
\urldef\tempurl%
\url{https://doi.org/10.1007/978-3-031-40113-8_9}
\showDOI{\tempurl}


\bibitem[Singh et~al\mbox{.}(2023)]%
        {10.1145/3579363}
\bibfield{author}{\bibinfo{person}{Ronal Singh}, \bibinfo{person}{Tim Miller}, \bibinfo{person}{Henrietta Lyons}, \bibinfo{person}{Liz Sonenberg}, \bibinfo{person}{Eduardo Velloso}, \bibinfo{person}{Frank Vetere}, \bibinfo{person}{Piers Howe}, {and} \bibinfo{person}{Paul Dourish}.} \bibinfo{year}{2023}\natexlab{}.
\newblock \showarticletitle{Directive Explanations for Actionable Explainability in Machine Learning Applications}.
\newblock  (\bibinfo{year}{2023}).
\newblock
\showISSN{2160-6455}
\urldef\tempurl%
\url{https://doi.org/10.1145/3579363}
\showDOI{\tempurl}


\bibitem[Smith(2018)]%
        {smith2018generalizability}
\bibfield{author}{\bibinfo{person}{Brett Smith}.} \bibinfo{year}{2018}\natexlab{}.
\newblock \showarticletitle{Generalizability in qualitative research: Misunderstandings, opportunities and recommendations for the sport and exercise sciences}.
\newblock \bibinfo{journal}{\emph{Qualitative research in sport, exercise and health}} \bibinfo{volume}{10}, \bibinfo{number}{1} (\bibinfo{year}{2018}), \bibinfo{pages}{137--149}.
\newblock
\urldef\tempurl%
\url{https://doi.org/10.1080/2159676X.2017.1393221}
\showDOI{\tempurl}


\bibitem[Spinner et~al\mbox{.}(2024)]%
        {10.1145/3652028}
\bibfield{author}{\bibinfo{person}{Thilo Spinner}, \bibinfo{person}{Rebecca Kehlbeck}, \bibinfo{person}{Rita Sevastjanova}, \bibinfo{person}{Tobias St\"{a}hle}, \bibinfo{person}{Daniel~A. Keim}, \bibinfo{person}{Oliver Deussen}, {and} \bibinfo{person}{Mennatallah El-Assady}.} \bibinfo{year}{2024}\natexlab{}.
\newblock \showarticletitle{generAItor: Tree-in-the-Loop Text Generation for Language Model Explainability and Adaptation}.
\newblock  (\bibinfo{year}{2024}).
\newblock
\showISSN{2160-6455}
\urldef\tempurl%
\url{https://doi.org/10.1145/3652028}
\showDOI{\tempurl}


\bibitem[Springer and Whittaker(2020)]%
        {10.1145/3374218}
\bibfield{author}{\bibinfo{person}{Aaron Springer} {and} \bibinfo{person}{Steve Whittaker}.} \bibinfo{year}{2020}\natexlab{}.
\newblock \showarticletitle{Progressive Disclosure: When, Why, and How Do Users Want Algorithmic Transparency Information?}
\newblock  (\bibinfo{year}{2020}).
\newblock
\showISSN{2160-6455}
\urldef\tempurl%
\url{https://doi.org/10.1145/3374218}
\showDOI{\tempurl}


\bibitem[Terry et~al\mbox{.}(2017)]%
        {terry2017thematic}
\bibfield{author}{\bibinfo{person}{Gareth Terry}, \bibinfo{person}{Nikki Hayfield}, \bibinfo{person}{Victoria Clarke}, {and} \bibinfo{person}{Virginia Braun}.} \bibinfo{year}{2017}\natexlab{}.
\newblock \showarticletitle{Thematic analysis}.
\newblock \bibinfo{journal}{\emph{The SAGE handbook of qualitative research in psychology}}  \bibinfo{volume}{2} (\bibinfo{year}{2017}), \bibinfo{pages}{17--37}.
\newblock
\urldef\tempurl%
\url{https://doi.org/10.4135/9781526405555.n2}
\showDOI{\tempurl}


\bibitem[Teubner et~al\mbox{.}(2023)]%
        {teubner2023welcome}
\bibfield{author}{\bibinfo{person}{Timm Teubner}, \bibinfo{person}{Christoph~M Flath}, \bibinfo{person}{Christof Weinhardt}, \bibinfo{person}{Wil van~der Aalst}, {and} \bibinfo{person}{Oliver Hinz}.} \bibinfo{year}{2023}\natexlab{}.
\newblock \showarticletitle{Welcome to the era of chatgpt et al. the prospects of large language models}.
\newblock \bibinfo{journal}{\emph{Business \& Information Systems Engineering}} \bibinfo{volume}{65}, \bibinfo{number}{2} (\bibinfo{year}{2023}), \bibinfo{pages}{95--101}.
\newblock
\urldef\tempurl%
\url{https://doi.org/10.1007/s12599-023-00795-x}
\showDOI{\tempurl}


\bibitem[Tian et~al\mbox{.}({[n.\,d.]})]%
        {tian2023Opportunities}
\bibfield{author}{\bibinfo{person}{Shubo Tian}, \bibinfo{person}{Qiao Jin}, \bibinfo{person}{Lana Yeganova}, \bibinfo{person}{Po-Ting Lai}, \bibinfo{person}{Qingqing Zhu}, \bibinfo{person}{Xiuying Chen}, \bibinfo{person}{Yifan Yang}, \bibinfo{person}{Qingyu Chen}, \bibinfo{person}{Won Kim}, \bibinfo{person}{Donald~C. Comeau}, \bibinfo{person}{Rezarta Islamaj}, \bibinfo{person}{Aadit Kapoor}, \bibinfo{person}{Xin Gao}, {and} \bibinfo{person}{Zhiyong Lu}.} \bibinfo{year}{[n.\,d.]}\natexlab{}.
\newblock \bibinfo{booktitle}{\emph{Opportunities and {{Challenges}} for {{ChatGPT}} and {{Large Language Models}} in {{Biomedicine}} and {{Health}}}}.
\newblock
\urldef\tempurl%
\url{https://doi.org/10.48550/arXiv.2306.10070}
\showDOI{\tempurl}
\showeprint[arxiv]{2306.10070}~[cs, q-bio]


\bibitem[Tolmeijer et~al\mbox{.}(2022)]%
        {10.1145/3491102.3517732}
\bibfield{author}{\bibinfo{person}{Suzanne Tolmeijer}, \bibinfo{person}{Markus Christen}, \bibinfo{person}{Serhiy Kandul}, \bibinfo{person}{Markus Kneer}, {and} \bibinfo{person}{Abraham Bernstein}.} \bibinfo{year}{2022}\natexlab{}.
\newblock \showarticletitle{Capable but Amoral? Comparing AI and Human Expert Collaboration in Ethical Decision Making}. In \bibinfo{booktitle}{\emph{Proceedings of the 2022 CHI Conference on Human Factors in Computing Systems}} (New Orleans, LA, USA) \emph{(\bibinfo{series}{CHI '22})}. \bibinfo{publisher}{Association for Computing Machinery}, \bibinfo{address}{New York, NY, USA}, Article \bibinfo{articleno}{160}, \bibinfo{numpages}{17}~pages.
\newblock
\showISBNx{9781450391573}
\urldef\tempurl%
\url{https://doi.org/10.1145/3491102.3517732}
\showDOI{\tempurl}


\bibitem[Vainio-Pekka et~al\mbox{.}(2023)]%
        {10.1145/3599974}
\bibfield{author}{\bibinfo{person}{Heidi Vainio-Pekka}, \bibinfo{person}{Mamia Ori-Otse Agbese}, \bibinfo{person}{Marianna Jantunen}, \bibinfo{person}{Ville Vakkuri}, \bibinfo{person}{Tommi Mikkonen}, \bibinfo{person}{Rebekah Rousi}, {and} \bibinfo{person}{Pekka Abrahamsson}.} \bibinfo{year}{2023}\natexlab{}.
\newblock \showarticletitle{The Role of Explainable AI in the Research Field of AI Ethics}.
\newblock  (\bibinfo{year}{2023}).
\newblock
\urldef\tempurl%
\url{https://doi.org/10.1145/3599974}
\showDOI{\tempurl}


\bibitem[Wang and Jin({[n.\,d.]})]%
        {wang2023Brief}
\bibfield{author}{\bibinfo{person}{Shuyue Wang} {and} \bibinfo{person}{Pan Jin}.} \bibinfo{year}{[n.\,d.]}\natexlab{}.
\newblock \showarticletitle{A {{Brief Summary}} of {{Prompting}} in {{Using GPT Models}}}.
\newblock  (\bibinfo{year}{[n.\,d.]}).
\newblock
\showISSN{2632-3834}
\urldef\tempurl%
\url{https://doi.org/10.32388/IMZI2Q}
\showDOI{\tempurl}


\bibitem[Wang and Yin(2022)]%
        {10.1145/3519266}
\bibfield{author}{\bibinfo{person}{Xinru Wang} {and} \bibinfo{person}{Ming Yin}.} \bibinfo{year}{2022}\natexlab{}.
\newblock \showarticletitle{Effects of Explanations in AI-Assisted Decision Making: Principles and Comparisons}.
\newblock  (\bibinfo{year}{2022}).
\newblock
\showISSN{2160-6455}
\urldef\tempurl%
\url{https://doi.org/10.1145/3519266}
\showDOI{\tempurl}


\bibitem[Watkins(2012)]%
        {watkins2012qualitative}
\bibfield{author}{\bibinfo{person}{Daphne~C Watkins}.} \bibinfo{year}{2012}\natexlab{}.
\newblock \showarticletitle{Qualitative research: The importance of conducting research that doesn’t “count”}.
\newblock \bibinfo{journal}{\emph{Health promotion practice}} \bibinfo{volume}{13}, \bibinfo{number}{2} (\bibinfo{year}{2012}), \bibinfo{pages}{153--158}.
\newblock
\urldef\tempurl%
\url{https://doi.org/10.1177/152483991243737}
\showDOI{\tempurl}


\bibitem[Wei et~al\mbox{.}({[n.\,d.]})]%
        {wei2022ChainofThought}
\bibfield{author}{\bibinfo{person}{Jason Wei}, \bibinfo{person}{Xuezhi Wang}, \bibinfo{person}{Dale Schuurmans}, \bibinfo{person}{Maarten Bosma}, \bibinfo{person}{Brian Ichter}, \bibinfo{person}{Fei Xia}, \bibinfo{person}{Ed Chi}, \bibinfo{person}{Quoc~V. Le}, {and} \bibinfo{person}{Denny Zhou}.} \bibinfo{year}{[n.\,d.]}\natexlab{}.
\newblock \showarticletitle{Chain-of-{{Thought Prompting Elicits Reasoning}} in {{Large Language Models}}}.
\newblock   \bibinfo{volume}{35} (\bibinfo{year}{[n.\,d.]}), \bibinfo{pages}{24824--24837}.
\newblock
\urldef\tempurl%
\url{https://proceedings.neurips.cc/paper_files/paper/2022/hash/9d5609613524ecf4f15af0f7b31abca4-Abstract-Conference.html}
\showURL{%
\tempurl}


\bibitem[White et~al\mbox{.}(2012)]%
        {white2012management}
\bibfield{author}{\bibinfo{person}{Debbie~Elizabeth White}, \bibinfo{person}{Nelly~D Oelke}, {and} \bibinfo{person}{Steven Friesen}.} \bibinfo{year}{2012}\natexlab{}.
\newblock \showarticletitle{Management of a large qualitative data set: Establishing trustworthiness of the data}.
\newblock \bibinfo{journal}{\emph{International journal of qualitative methods}} \bibinfo{volume}{11}, \bibinfo{number}{3} (\bibinfo{year}{2012}), \bibinfo{pages}{244--258}.
\newblock
\urldef\tempurl%
\url{https://doi.org/10.1177/160940691201100305}
\showDOI{\tempurl}


\bibitem[White et~al\mbox{.}(2023)]%
        {white2023prompt}
\bibfield{author}{\bibinfo{person}{Jules White}, \bibinfo{person}{Quchen Fu}, \bibinfo{person}{Sam Hays}, \bibinfo{person}{Michael Sandborn}, \bibinfo{person}{Carlos Olea}, \bibinfo{person}{Henry Gilbert}, \bibinfo{person}{Ashraf Elnashar}, \bibinfo{person}{Jesse Spencer-Smith}, {and} \bibinfo{person}{Douglas~C. Schmidt}.} \bibinfo{year}{2023}\natexlab{}.
\newblock \bibinfo{title}{A Prompt Pattern Catalog to Enhance Prompt Engineering with ChatGPT}.
\newblock
\newblock
\showeprint[arxiv]{2302.11382}~[cs.SE]


\bibitem[Williams and Moser(2019)]%
        {williams2019art}
\bibfield{author}{\bibinfo{person}{Michael Williams} {and} \bibinfo{person}{Tami Moser}.} \bibinfo{year}{2019}\natexlab{}.
\newblock \showarticletitle{The art of coding and thematic exploration in qualitative research}.
\newblock \bibinfo{journal}{\emph{International Management Review}} \bibinfo{volume}{15}, \bibinfo{number}{1} (\bibinfo{year}{2019}), \bibinfo{pages}{45--55}.
\newblock
\showISSN{15516849}
\urldef\tempurl%
\url{http://www.imrjournal.org/uploads/1/4/2/8/14286482/imr-v15n1art4.pdf}
\showURL{%
\tempurl}


\bibitem[Wu et~al\mbox{.}(2023)]%
        {10113601}
\bibfield{author}{\bibinfo{person}{Tianyu Wu}, \bibinfo{person}{Shizhu He}, \bibinfo{person}{Jingping Liu}, \bibinfo{person}{Siqi Sun}, \bibinfo{person}{Kang Liu}, \bibinfo{person}{Qing-Long Han}, {and} \bibinfo{person}{Yang Tang}.} \bibinfo{year}{2023}\natexlab{}.
\newblock \showarticletitle{A Brief Overview of ChatGPT: The History, Status Quo and Potential Future Development}.
\newblock \bibinfo{journal}{\emph{IEEE/CAA Journal of Automatica Sinica}} \bibinfo{volume}{10}, \bibinfo{number}{5} (\bibinfo{year}{2023}), \bibinfo{pages}{1122--1136}.
\newblock
\urldef\tempurl%
\url{https://doi.org/10.1109/JAS.2023.123618}
\showDOI{\tempurl}


\bibitem[Yuan et~al\mbox{.}(2023)]%
        {yuan2023no}
\bibfield{author}{\bibinfo{person}{Zhiqiang Yuan}, \bibinfo{person}{Yiling Lou}, \bibinfo{person}{Mingwei Liu}, \bibinfo{person}{Shiji Ding}, \bibinfo{person}{Kaixin Wang}, \bibinfo{person}{Yixuan Chen}, {and} \bibinfo{person}{Xin Peng}.} \bibinfo{year}{2023}\natexlab{}.
\newblock \bibinfo{title}{No More Manual Tests? Evaluating and Improving ChatGPT for Unit Test Generation}.
\newblock
\newblock
\showeprint[arxiv]{2305.04207}~[cs.SE]


\bibitem[Yue et~al\mbox{.}({[n.\,d.]})]%
        {yue2019survey}
\bibfield{author}{\bibinfo{person}{Lin Yue}, \bibinfo{person}{Weitong Chen}, \bibinfo{person}{Xue Li}, \bibinfo{person}{Wanli Zuo}, {and} \bibinfo{person}{Minghao Yin}.} \bibinfo{year}{[n.\,d.]}\natexlab{}.
\newblock \showarticletitle{A Survey of Sentiment Analysis in Social Media}.
\newblock  \bibinfo{volume}{60}, \bibinfo{number}{2} (\bibinfo{year}{[n.\,d.]}), \bibinfo{pages}{617--663}.
\newblock
\showISSN{0219-3116}
\urldef\tempurl%
\url{https://doi.org/10.1007/s10115-018-1236-4}
\showDOI{\tempurl}


\bibitem[Zamfirescu-Pereira et~al\mbox{.}(2023a)]%
        {10.1145/3563657.3596138}
\bibfield{author}{\bibinfo{person}{J.D. Zamfirescu-Pereira}, \bibinfo{person}{Heather Wei}, \bibinfo{person}{Amy Xiao}, \bibinfo{person}{Kitty Gu}, \bibinfo{person}{Grace Jung}, \bibinfo{person}{Matthew~G Lee}, \bibinfo{person}{Bjoern Hartmann}, {and} \bibinfo{person}{Qian Yang}.} \bibinfo{year}{2023}\natexlab{a}.
\newblock \showarticletitle{Herding AI Cats: Lessons from Designing a Chatbot by Prompting GPT-3}. In \bibinfo{booktitle}{\emph{Proceedings of the 2023 ACM Designing Interactive Systems Conference}} (Pittsburgh, PA, USA) \emph{(\bibinfo{series}{DIS '23})}. \bibinfo{publisher}{Association for Computing Machinery}, \bibinfo{address}{New York, NY, USA}, \bibinfo{pages}{2206–2220}.
\newblock
\showISBNx{9781450398930}
\urldef\tempurl%
\url{https://doi.org/10.1145/3563657.3596138}
\showDOI{\tempurl}


\bibitem[Zamfirescu-Pereira et~al\mbox{.}(2023b)]%
        {10.1145/3544548.3581388}
\bibfield{author}{\bibinfo{person}{J.D. Zamfirescu-Pereira}, \bibinfo{person}{Richmond~Y. Wong}, \bibinfo{person}{Bjoern Hartmann}, {and} \bibinfo{person}{Qian Yang}.} \bibinfo{year}{2023}\natexlab{b}.
\newblock \showarticletitle{Why Johnny Can’t Prompt: How Non-AI Experts Try (and Fail) to Design LLM Prompts}. In \bibinfo{booktitle}{\emph{Proceedings of the 2023 CHI Conference on Human Factors in Computing Systems}} (Hamburg, Germany) \emph{(\bibinfo{series}{CHI '23})}. \bibinfo{publisher}{Association for Computing Machinery}, \bibinfo{address}{New York, NY, USA}, Article \bibinfo{articleno}{437}, \bibinfo{numpages}{21}~pages.
\newblock
\showISBNx{9781450394215}
\urldef\tempurl%
\url{https://doi.org/10.1145/3544548.3581388}
\showDOI{\tempurl}


\bibitem[Zerilli et~al\mbox{.}(2022)]%
        {zerilli2022transparency}
\bibfield{author}{\bibinfo{person}{John Zerilli}, \bibinfo{person}{Umang Bhatt}, {and} \bibinfo{person}{Adrian Weller}.} \bibinfo{year}{2022}\natexlab{}.
\newblock \showarticletitle{How transparency modulates trust in artificial intelligence}.
\newblock \bibinfo{journal}{\emph{Patterns}} (\bibinfo{year}{2022}).
\newblock
\urldef\tempurl%
\url{https://doi.org/10.1016/j.patter.2022.100455}
\showDOI{\tempurl}


\bibitem[Zhao et~al\mbox{.}({[n.\,d.]})]%
        {zhao2021Calibrate}
\bibfield{author}{\bibinfo{person}{Zihao Zhao}, \bibinfo{person}{Eric Wallace}, \bibinfo{person}{Shi Feng}, \bibinfo{person}{Dan Klein}, {and} \bibinfo{person}{Sameer Singh}.} \bibinfo{year}{[n.\,d.]}\natexlab{}.
\newblock \showarticletitle{Calibrate {{Before Use}}: {{Improving Few-shot Performance}} of {{Language Models}}}. In \bibinfo{booktitle}{\emph{Proceedings of the 38th {{International Conference}} on {{Machine Learning}}}} (2021-07-01). \bibinfo{publisher}{{PMLR}}, \bibinfo{pages}{12697--12706}.
\newblock
\showISSN{2640-3498}
\urldef\tempurl%
\url{https://proceedings.mlr.press/v139/zhao21c.html}
\showURL{%
\tempurl}


\end{thebibliography}

\appendix
\section{Appendix}

\end{document}